\documentclass[11pt]{article}

\newcommand{\be}{\begin{equation}}
\newcommand{\ee}{\end{equation}}
\newcommand{\bea}{\begin{eqnarray}}
\newcommand{\eea}{\end{eqnarray}}
\newcommand{\sn}{{\rm sn}}

\newcommand{\dn}{{\rm dn}}
\newcommand{\cn}{{\rm cn}}
\newcommand{\sech}{{\rm sech}}

\begin{document}
\vspace{.5in}
\begin{center}
{\LARGE{\bf Domain Wall and Periodic Solutions of a Coupled $\phi^6$ Model}}
\end{center}

\vspace{.3in}
\begin{center}
{\LARGE{\bf Avinash Khare}} \\
{Institute for Physics, Bhubaneswar, Orissa 751005, India}
\end{center}

\begin{center}
{\LARGE{\bf Avadh Saxena}} \\
{Theoretical Division and Center for Nonlinear Studies, Los
Alamos National Lab, Los Alamos, NM 87545, USA}
\end{center}

\vspace{.9in}
{\bf {Abstract:}}

Coupled triple well ($\phi^6$) one-dimensional potentials occur in
both condensed matter physics and field theory.  Here we provide a
set of exact periodic solutions in terms of elliptic functions (domain 
wall arrays) and obtain single domain wall solutions in specific 
limits.  Both topological, nontopological (e.g. some pulse-like 
solutions) as well as mixed domain walls are obtained.  We relate 
these solutions to structural phase transitions in materials with 
polarization, shuffle modes and strain.  We calculate the energy and 
the asymptotic interaction between solitons for various solutions.  
We also consider the discrete analog of these coupled models and 
obtain several single and periodic domain wall exact solutions.

\newpage

\section{Introduction}

Many physical problems of interest often have a coupling between two 
order parameters.  In a recent paper we obtained exact single and 
periodic domain wall solutions of coupled double well ($\phi^4$) models 
\cite{ks} (both continuum and discrete) which arise in many second 
order transitions \cite{aubry,abel,jcp}.  Similarly, coupled triple 
well or $\phi^6$ models \cite{sbh} arise in the context of first order 
phase transitions.  There exist analogous coupled models in field 
theoretical contexts \cite{bazeia,santos,lou}.  Specifically, when a 
first order transition is driven by two primary order parameters, the 
free energy should be expanded to sixth order in both order parameters 
with a biquadratic (or possibly other symmetry allowed) coupling.  An 
example of such a transition is the cubic to monoclinic structural 
transformation in shape memory alloys NiTi and AuCd in which two phonon 
(or shuffle) modes are simultaneously active \cite{hs}.  A simpler 
example is the square to oblique lattice transition driven simultaneously 
by two different strain components with a biquadratic coupling \cite{hlss}.  
Here our motivation is to obtain all possible single and periodic 
domain wall solutions of these models and then connect to experimental 
observations wherever possible.

The paper is organized as follows.  In Sec. II we provide {\it twenty}  
distinct periodic solutions of a coupled continuum $\phi^6$ model.  In 
doing so, we obtain previously unknown periodic solutions of the standard 
(uncoupled) $\phi^6$ model \cite{behera,falk,sanati}.  We also calculate 
the total energy and the interaction energy between solitons for all the 
solutions.  Section III contains {\it six} different solutions for the 
corresponding discrete $\phi^6$ case.  The latter arises in the context 
of first order phase transitions on a lattice.  Finally, we conclude 
in Sec. IV with remarks on related coupled models.

\section{Solutions of a Coupled $\phi^6$ Continuum Model}

In the literature, the examples of a coupled $\phi^6$ model are scarce.  
The reason is presumably that one field (or order parameter) is usually  
dominant and leads to a first order transition.  One does not need
nonlinearity in the secondary order parameter (or the field) to drive
the transition.  However, in crystalline phase transitions this is not
necessarily true and within the context of Landau theory, in certain 
transitions symmetry allows one to go to the sixth order nonlinearity 
in both fields (as primary order parameters) with a biquadratic coupling 
\cite{sbh,hs,hlss}.  In this paper, we adopt this viewpoint and consider 
the solutions of a coupled $\phi^6$ model with the potential
\be\label{2.1}
V(\phi,\psi)=\left(\frac{a_1}{2} \phi^2 -\frac{b_1}{4}\phi^4 
+\frac{c_1}{6}\phi^6\right)  
+\left(\frac{a_2}{2} \psi^2 -\frac{b_2}{4}\psi^4 +\frac{c_2}{6}\psi^6 
\right) +\frac{d}{2}\phi^2 \psi^2\,, 
\ee
where $a_{1,2},b_{1,2},c_{1,2}$ and $d$ are material (or system) dependent
parameters; $\phi$ and $\psi$ are scalar fields.  A comprehensive analysis 
of this coupled continuum $\phi^6$ model shows that it has {\it twenty} 
possible periodic solutions, i.e. eleven ``bright-bright", three 
``bright-dark" and six ``dark-dark" solutions. In particular, we obtain 
six solutions below the transition temperature $T_c$, ten at $T_c$, one 
above $T_c$ and three in mixed phase in the sense that while one of the 
field is above $T_c$, the other one is below $T_c$.  The latter situation 
is akin to the one found in multiferroic materials where one transition 
(e.g. antiferromagnetic) takes place at a higher temperature than the 
other transition (e.g. ferroelectric) or vice versa \cite{kimura,fiebig}.  
However, in the corresponding discrete case we are able to obtain only 
{\it six} periodic solutions. 

The (static) equations of motion which follow from Eq. (\ref{2.1}) are
\bea\label{2.2}
&&\frac{d^2 \phi}{dx^2}= a_1 \phi -b_1\phi^3 +c_1 \phi^5 +d\phi \psi^2\,, 
\nonumber \\
&&\frac{d^2 \psi}{dx^2}= a_2 \psi -b_2\psi^3 +c_2 \psi^5 +d\psi \phi^2\,. 
\eea
These coupled equations have twenty distinct periodic (elliptic function) 
solutions which we discuss one by one systematically. It is worth pointing 
out that in turn in the single soliton limit, these lead to {\it nine} 
distinct (hyperbolic) soliton solutions.  In particular, one obtains three 
solutions below the transition temperature $T_c$, three at $T_c$, one above 
$T_c$ and two in the mixed phase in the sense that while one of the field 
is above $T_c$, the other one is below $T_c$.

For static solutions the energy is given by
\be\label{2.3}
E=\int \bigg [\frac{1}{2}\left(\frac{d\phi}{dx}\right)^2+\frac{1}{2} 
\left(\frac{d\psi}{dx}\right)^2 +V(\phi,\psi) \bigg ]\,dx\,,
\ee
where the limits of integration are from $-\infty$ to $\infty$ for 
hyperbolic solutions (single solitons) on the full line. On the other
hand, for the periodic solutions (i.e. soliton lattices), the limits are
from $-K(k)$ to $K(k)$. Here $K(k)$ (and $E(k)$ below) denote complete 
elliptic integral of the first (and second) kind \cite{gr}.

Using the equations of motion, one can show that for all of our solutions 
\be\label{2.4}
V(\phi,\psi)= \bigg [\frac{1}{2} \left(\frac{d\phi}{dx}\right)^2
+\frac{1}{2}\left(\frac{d\psi}{dx}\right)^2 \bigg ] +C\,,
\ee
where the constant $C$ in general varies from solution to solution.
Hence the energy measured with respect to the reference energy (e.g. 
the local or global minimum) $\hat{E}=E-\int C\,dx$ is given by
\be\label{2.5}
\hat{E} \equiv E-\int C\,dx = \int \bigg [\left(\frac{d\phi}{dx}\right)^2 
+\left(\frac{d\psi}{dx}\right)^2 \bigg ]\,dx\,.
\ee
Below we give an explicit expression for the energy in the case of all 
twenty periodic solutions (and hence the corresponding nine hyperbolic 
solutions). In each case we also give an expression for the constant 
$C$.  To begin with, we first obtain ten bright-bright solutions at $T_c$.

\subsection{Solutions I and II}

We look for the most general solutions to the coupled Eqs. (\ref{2.2}) 
in terms of Jacobi elliptic functions $\sn(x,m)$, $\cn(x,m)$ and $\dn(x,m)$ 
\cite{gr} where the modulus $m \equiv k^2$.  It is easily shown that
\be\label{2.6}
\phi=A\sqrt{1\pm\sn(Bx+x_0,m)}\,,~~\psi=F\sqrt{1\pm\sn(Bx+x_0,m)}\,,
\ee
are solutions to the coupled Eqs. (\ref{2.2}) 
provided the following six coupled equations are satisfied
\be\label{11}
a_1-b_1 A^2+dF^2+c_1 A^4 = \frac{-B^2}{4}\,,
\ee
\be\label{12}
-b_1 A^2+dF^2+2c_1 A^4 = \frac{-mB^2}{2}\,,
\ee
\be\label{13}
c_1 A^4 = \frac{3mB^2}{4}\,,
\ee
\be\label{14}
a_2-b_2 F^2+dA^2+c_2 F^4 = \frac{-B^2}{4}\,,
\ee
\be\label{15}
-b_2 F^2+dA^2+2c_2 F^4 = \frac{-mB^2}{2}\,,
\ee
\be\label{16}
c_2 F^4 = \frac{3mB^2}{4}\,.
\ee
Here $+ (-)$ sign in $\phi$ goes with $+ (-)$ sign in $\psi$, 
$A$ and $F$ denote the amplitudes of the kink lattice, $B$ is an
inverse characteristic length while $x_0$ is the (arbitrary) location of
the kink. Three of these equations determine the three unknowns $A,F,B$
while the other three equations give three constraints between the seven
parameters $a_{1,2},b_{1,2},c_{1,2},d$. In particular, $A,F,B$ are given
by
\be\label{2.7}
B^2=\frac{4a_1}{(5m-1)}\,,
~~A^2=\frac{8m(d+b_2)a_1}{(5m-1)(b_1b_2-d^2)}\,, 
~~F^2=\frac{8m(d+b_1)a_1}{(5m-1)(b_1b_2-d^2)}\,, 
\ee
while the three constraints are
\be\label{2.8}
a_1=a_2\,,~~c_1(d+b_2)^2=c_2(d+b_1)^2\,,~~
(b_1b_2-d^2)^2 =\frac{64ma_1c_1(d+b_2)^2}{3(5m-1)}\,.
\ee
For $d=0$ the last constraint reduces to the constraint (for the uncoupled 
$\phi^6$ \cite{behera,falk,sanati}  
model)  
\be\label{2.9}
b_1^2 = \frac{64ma_1c_1}{3(5m-1)}\,.
\ee

So far as we are aware of, the uncoupled (i.e. $d=0$) kink lattice solution
$\phi=A\sqrt{1\pm \sn(Bx+x_0,m)}$,  satisfying
\be
B^2=\frac{4a_1}{(5m-1)}\,,
~~A^2=\frac{8ma_1}{(5m-1)b_1}\,,
\ee
and the constraint (\ref{2.9}) have never been explicitly written down
before in the literature.

At $m=1$ these solutions go over to the
topological (bright-bright) solutions
\be\label{2.10}
\phi=A\sqrt{1\pm\tanh(Bx+x_0)}\,,~~\psi=F\sqrt{1\pm\tanh(Bx+x_0)}\,,
\ee
provided
\be\label{2.11}
B^2=a_1\,,
~~A^2=\frac{2(d+b_2)a_1}{(b_1b_2-d^2)}\,, 
~~F^2=\frac{2(d+b_1)a_1}{(b_1b_2-d^2)}\,. 
\ee
Further, two of the constraints remain the same as above [Eq. (\ref{2.8})]  
while the third one is now given by
\be\label{2.12}
(b_1b_2-d^2)^2 =(16/3)a_1c_1(d+b_2)^2\,,
\ee
which for $d=0$ reduces to  
\be\label{2.13}
b_1^2=(16/3)a_1c_1\,. 
\ee
This corresponds to the point (i.e. temperature) where there are three 
degenerate minima (or the point where a discontinuous transition takes 
place) of the usual (uncoupled) $\phi^6$ field theory 
\cite{behera,falk,sanati}.

{\bf Energy}: Corresponding to the periodic solutions (\ref{2.6}), 
the energy $\hat{E}$ and the constant $C$ are given by
\bea\label{2.14}
&&\hat{E}=\frac{(A^2+F^2)B}{4}E(k)\,, \nonumber \\
&&C=-\frac{1}{4}(1-k^2)(A^2+F^2)B^2\,.
\eea
On using the expansion formulas for $E(k)$ [and $K(k)$ to be used 
later in the paper] around $k=1$ 
(note $m=k^2$) as given in \cite{gr}
\be\label{2.15}
E(k)=1+\frac{k'^2}{2}\left[\ln\left(\frac{4}{k'}\right)-\frac{1}{2} 
\right]+...\,, 
\ee
\be\label{2.16}
K(k)=\ln\left(\frac{4}{k'}\right)+\frac{k'^2}{4}\left[\ln\left( 
\frac{4}{k'}\right)-1\right]+...\,,~~k'^2=1-k^2\,,
\ee
the energy of the periodic solution can be rewritten 
as the energy of the corresponding hyperbolic (bright-bright) soliton 
solution [Eq. (\ref{2.10})]  
plus the asymptotic (i.e. widely separated solitons) interaction energy 
\cite{sanati}. We find
\be\label{2.17}
\hat{E}=E_{kink}+E_{int}=(A^2+F^2)B \left(\frac{1}{4}
+\frac{k'^2}{8}\left[\ln\left(\frac{4}{k'}\right)-\frac{1}{2} \right] 
\right )\,. 
\ee
Note that these solutions exist only when $b_1 b_2 > d^2$.  The 
interaction energy vanishes at $k^2=1$, as it should.  

\subsection{Solution III}

It turns out that the coupled Eqs. (\ref{2.2}) also admit a novel mixed
solution at $T=T_c$ given by
\be\label{y.6}
\phi=A\sqrt{1+\sn(Bx+x_0,m)}\,,~~\psi=F\sqrt{1-\sn(Bx+x_0,m)}\,,
\ee
provided the following six coupled equations are satisfied
\be\label{y11}
a_1-b_1 A^2+dF^2+c_1 A^4 = \frac{-B^2}{4}\,,
\ee
\be\label{y12}
-b_1 A^2-dF^2+2c_1 A^4 = \frac{-mB^2}{2}\,,
\ee
\be\label{y13}
c_1 A^4 = \frac{3mB^2}{4}\,,
\ee
\be\label{y14}
a_2-b_2 F^2+dA^2+c_2 F^4 = \frac{-B^2}{4}\,,
\ee
\be\label{y15}
-b_2 F^2-dA^2+2c_2 F^4 = \frac{-mB^2}{2}\,,
\ee
\be\label{y16}
c_2 F^4 = \frac{3mB^2}{4}\,.
\ee
Three of these equations determine the three unknowns $A,F,B$
while the other three equations give three constraints between the seven
parameters $a_{1,2},b_{1,2},c_{1,2},d$. In particular, $A,F,B$ are given
by
\bea\label{y.7}
&&B^2=\frac{4a_1(b_1 b_2-d^2)}{[(5m-1)b_1 b_2-16mb_1 d
+(11m+1)d^2)]}\,, \nonumber \\
&&A^2=\frac{8ma_1(b_2-d)}{[(5m-1)b_1 b_2-16mb_1 d+(11m+1)d^2)]}\,,
~~F^2=\frac{8ma_1(b_1-d)}{[(5m-1)b_1 b_2-16mb_1 d+(11m+1)d^2)]}\,,
\eea
while the three constraints are
\bea\label{y.8}
&&[(5m-1)b_1 b_2+(11m+1)d^2-16mb_2d]a_1=
[(5m-1)b_1 b_2+(11m+1)d^2-16mb_1d]a_2\,, \nonumber \\
&&c_1(b_2-d)^2=c_2(b_1-d)^2\,,~~
(b_1b_2-d^2) =\frac{64ma_1c_1(b_2-d)^2}
{3[(5m-1)b_1 b_2+(11m+1)d^2-16mb_1d]}\,.
\eea
For $d=0$ the last constraint reduces to the constraint (for the 
uncoupled $\phi^6$ model) as given by Eq. (\ref{2.9}).  Note that no 
new solution is obtained by having the $-\sn$ term inside the square 
root in $\phi$ and the $+\sn$ term inside the square root in $\psi$ 
in Eq. (\ref{y.6}) since such a solution is trivially obtained from 
Eq. (\ref{y.6}) by merely interchanging $A$ and $F$.  This is so 
because the potential $V(\phi,\psi)$ is completely symmetric in $\phi$ 
and $\psi$.  Similar comments apply to many of the other asymmetric 
solutions given below.  

At $m=1$ this solution goes over to the
topological (bright-bright) solution
\be\label{y.10}
\phi=A\sqrt{1+\tanh(Bx+x_0)}\,,~~\psi=F\sqrt{1-\tanh(Bx+x_0)}\,,
\ee
provided
\bea\label{y.11}
B^2=\frac{a_1(b_1 b_2-d^2)}{(b_1 b_2-b_1 d
+d^2)}\,, 
~~A^2=\frac{a_1(b_2-d)}{2(b_1 b_2-4b_1 d+3d^2)}\,,
~~F^2=\frac{a_1(b_1-d)}{2(b_1 b_2-4b_1 d+3d^2)}\,,
\eea
while the three constraints are
\bea\label{y.12}
&&\big (b_1 b_2+3d^2-4b_2d \big )a_1=
\big (b_1 b_2+3d^2-4b_1d \big )a_2\,, \nonumber \\
&&c_1(b_2-d)^2=c_2(b_1-d)^2\,,~~
(b_1b_2-d^2) =\frac{16a_1c_1(b_2-d)^2}
{3(b_1 b_2+3d^2-4b_1d)}\,.
\eea
For $d=0$, the last constraint reduces to the constraint (\ref{2.13})  
which is the point where there are three degenerate minima (i.e. the
point where a discontinuous transition takes place) of the usual
(uncoupled) $\phi^6$ field theory \cite{behera,falk,sanati}.

{\bf Energy}: Remarkably, corresponding to the periodic solution 
(\ref{y.6}), the energy $\hat{E}$ is again the same as that for solutions 
I and II and is given by Eq. (\ref{2.14}). Thus the interaction energy 
too remains unchanged and is again given by Eq.  (\ref{2.17}).  However, 
$C$ is different for this solution than those for solutions I and II and 
is given by 
\be\label{y.13}
C=\frac{(a_1A^2+a_2F^2)}{4}-\frac{B^2}{16}(A^2+F^2)(3+m)\,.
\ee
Note that this solution also exists only when $b_1 b_2 > d^2$.

\subsection{Solutions IV and V}

It is easily shown that
\be\label{z.6}
\phi=A\sqrt{1\pm \sqrt{m}\sn(Bx+x_0,m)}\,,
~~\psi=F\sqrt{1\pm \sqrt{m}\sn(Bx+x_0,m)}\,,
\ee
are also solutions to the coupled Eqs. (\ref{2.2}) 
provided the following six coupled equations are satisfied
\be\label{z11}
a_1-b_1 A^2+dF^2+c_1 A^4 = \frac{-mB^2}{4}\,,
\ee
\be\label{z12}
-b_1 A^2+dF^2+2c_1 A^4 = \frac{-B^2}{2}\,,
\ee
\be\label{z13}
c_1 A^4 = \frac{3B^2}{4}\,,
\ee
\be\label{z14}
a_2-b_2 F^2+dA^2+c_2 F^4 = \frac{-mB^2}{4}\,,
\ee
\be\label{z15}
-b_2 F^2+dA^2+2c_2 F^4 = \frac{-B^2}{2}\,,
\ee
\be\label{z16}
c_2 F^4 = \frac{3B^2}{4}\,.
\ee
Three of these equations determine the three unknowns $A,F,B$
while the other three equations give three constraints between the seven
parameters $a_{1,2},b_{1,2},c_{1,2},d$. In particular, $A,F,B$ are given
by
\be\label{z.7}
B^2=\frac{4a_1}{(5-m)}\,,
~~A^2=\frac{8(d+b_2)a_1}{(5-m)(b_1b_2-d^2)}\,, 
~~F^2=\frac{8(d+b_1)a_1}{(5-m)(b_1b_2-d^2)}\,, 
\ee
while the three constraints are
\be\label{z.8}
a_1=a_2\,,~~c_1(d+b_2)^2=c_2(d+b_1)^2\,,~~
(b_1b_2-d^2)^2 =\frac{64a_1c_1(d+b_2)^2}{3(5-m)}\,.
\ee
For $d=0$ the last one reduces to the constraint (for the uncoupled
$\phi^6$ model) \cite{behera,falk,sanati} 
\be\label{z.9}
b_1^2=\frac{64a_1c_1}{3(5-m)}\,.
\ee

So far as we are aware of, the uncoupled (i.e. $d=0$) kink lattice solution
$\phi=A\sqrt{1\pm \sqrt{m}\sn(Bx+x_0,m)}$,  satisfying
\be
B^2=\frac{4a_1}{(5-m)}\,,
~~A^2=\frac{8a_1}{(5-m)b_1}\,,
\ee
and the constraint (\ref{z.9}) have never been explicitly written down
before in the literature.

At $m=1$ these solutions go over to the topological (bright-bright) 
solutions (\ref{2.10}) satisfying the constraints (\ref{2.11}) to 
(\ref{2.13}).

{\bf Energy}: Corresponding to the periodic solutions (\ref{z.6}), 
the energy $\hat{E}$ and the constant $C$ are given by
\bea\label{z.14}
&&\hat{E}=\frac{(A^2+F^2)B}{4}[E(k)-k'^2K(k)]\,, \nonumber \\
&&C=\frac{1}{4}(1-k^2)(A^2+F^2)B^2\,.
\eea
On using the expansion formulas for $E(k)$ and $K(k)$ around $k=1$ 
(note $m=k^2$) as given by Eqs. (\ref{2.15}) and (\ref{2.16}), 
the energy of the periodic solutions can be rewritten 
as the energy of the corresponding hyperbolic (bright-bright) soliton 
solutions [Eq. (\ref{2.10})]  
plus the interaction energy. We find
\be\label{z.17}
\hat{E}=E_{kink}+E_{int}=(A^2+F^2)B \left(\frac{1}{4}
-\frac{k'^2}{8}\left[\ln\left(\frac{4}{k'}\right)+\frac{1}{2} \right] 
\right )\,. 
\ee
Note that these solutions also exist only when $b_1 b_2 > d^2$.  The 
interaction energy vanishes at $k=1$. 

\subsection{Solution VI}

It turns out that the coupled Eqs. (\ref{2.2}) also admit a novel mixed
solution at $T=T_c$ given by
\be\label{w.6}
\phi=A\sqrt{1+\sqrt{m}\sn(Bx+x_0,m)}\,,
~~\psi=F\sqrt{1-\sqrt{m}\sn(Bx+x_0,m)}\,,
\ee
provided the following six coupled equations are satisfied
\be\label{w11}
a_1-b_1 A^2+dF^2+c_1 A^4 = \frac{-mB^2}{4}\,,
\ee
\be\label{w12}
-b_1 A^2-dF^2+2c_1 A^4 = \frac{-B^2}{2}\,,
\ee
\be\label{w13}
c_1 A^4 = \frac{3B^2}{4}\,,
\ee
\be\label{w14}
a_2-b_2 F^2+dA^2+c_2 F^4 = \frac{-mB^2}{4}\,,
\ee
\be\label{w15}
-b_2 F^2-dA^2+2c_2 F^4 = \frac{-B^2}{2}\,,
\ee
\be\label{w16}
c_2 F^4 = \frac{3B^2}{4}\,.
\ee
Three of these equations determine the three unknowns $A,F,B$
while the other three equations give three constraints between the seven
parameters $a_{1,2},b_{1,2},c_{1,2},d$. In particular, $A,F,B$ are given
by
\bea\label{w.7}
&&B^2=\frac{4a_1(b_1 b_2-d^2)}{[(5-m)b_1 b_2-16b_1 d
+(11+m)d^2)]}\,, \nonumber \\
&&A^2=\frac{8ma_1(b_2-d)}{[(5-m)b_1 b_2-16b_1 d+(11+m)d^2)]}\,,
~~F^2=\frac{8ma_1(b_1-d)}{[(5-m)b_1 b_2-16b_1 d+(11+m)d^2)]}\,,
\eea
while the three constraints are
\bea\label{w.8}
&&[(5-m)b_1 b_2+(11+m)d^2-16b_2d]a_1=
[(5-m)b_1 b_2+(11+m)d^2-16b_1d]a_2\,, \nonumber \\
&&c_1(b_2-d)^2=c_2(b_1-d)^2\,,~~
(b_1b_2-d^2) =\frac{64a_1c_1(b_2-d)^2}
{3[(5-m)b_1 b_2+(11+m)d^2-16b_1d]}\,.
\eea
For $d=0$ the last one reduces to the constraint (\ref{z.9}).

At $m=1$ this solution goes over to the
topological (bright-bright) solution (\ref{y.10})
satisfying the same constraints as given by Eqs. (\ref{y.11}) and
(\ref{y.12}).

{\bf Energy}: Remarkably, corresponding to the periodic solution 
(\ref{w.6}), the energy $\hat{E}$ is again the same as that for solutions 
IV and V and is given by Eq. (\ref{z.14}). Thus the interaction energy 
too remains unchanged and is again given by Eq. (\ref{z.17}).  However, 
$C$ is different for this solution than those for solutions IV and V and 
is given by
\be\label{w.13}
C=\frac{(a_1A^2+a_2F^2)}{4}-\frac{B^2}{16}(A^2+F^2)(1+3m)\,.
\ee
Note that this solution also exists only when $b_1 b_2 > d^2$.

\subsection{Solutions VII and VIII}

Apart from the above six solutions, at $T=T_c$, the field equations also
admit mixed type of solutions. In particular, it is easily shown that
\be\label{v.6}
\phi=A\sqrt{1\pm\sn(Bx+x_0,m)}\,,~~\psi=F\sqrt{1\pm \sqrt{m}\sn(Bx+x_0,m)}\,,
\ee
are solutions to the coupled Eqs. (\ref{2.2}) 
provided the following six coupled equations are satisfied
\be\label{v11}
a_1-b_1 A^2+dF^2+c_1 A^4 = \frac{-B^2}{4}\,,
\ee
\be\label{v12}
-b_1 A^2+\sqrt{m}dF^2+2c_1 A^4 = \frac{-mB^2}{2}\,,
\ee
\be\label{v13}
c_1 A^4 = \frac{3mB^2}{4}\,,
\ee
\be\label{v14}
a_2-b_2 F^2+dA^2+c_2 F^4 = \frac{-mB^2}{4}\,,
\ee
\be\label{v15}
-\sqrt{m}b_2 F^2+dA^2+2\sqrt{m}c_2 F^4 = \frac{-\sqrt{m}B^2}{2}\,,
\ee
\be\label{v16}
c_2 F^4 = \frac{3B^2}{4}\,.
\ee
Three of these equations determine the three unknowns $A,F,B$
while the other three equations give three constraints between the seven
parameters $a_{1,2},b_{1,2},c_{1,2},d$. In particular, $A,F,B$ are given
by
\bea\label{v.7}
&&B^2=\frac{4a_1(b_1 b_2-d^2)}{[(5m-1)b_1 b_2-(8\sqrt{m}-3m-1)d^2-
8db_1(1-\sqrt{m})]}\,, \nonumber \\
&&A^2=\frac{8\sqrt{m}(d+\sqrt{m}b_2)a_1}
{[(5m-1)b_1 b_2-(8\sqrt{m}-3m-1)d^2- 8db_1(1-\sqrt{m})]}\,, \nonumber \\
&&F^2=\frac{8(b_1+\sqrt{m}d)a_1}
{[(5m-1)b_1 b_2-(8\sqrt{m}-3m-1)d^2- 8db_1(1-\sqrt{m})]}\,, 
\eea
while the three constraints are
\bea\label{v.8}
&&[(5m-1)b_1 b_2-(8\sqrt{m}-3m-1)d^2- 8db_1(1-\sqrt{m})]a_2
\nonumber \\ 
&&=[(5-m)b_1 b_2-(8\sqrt{m}-3-m)d^2
+ 8db_2\sqrt{m}(1-\sqrt{m})]a_1\,,~~
c_1(d+\sqrt{m}b_2)^2=c_2(b_1+\sqrt{m}d)^2\,, \nonumber \\
&&(b_1b_2-d^2) =\frac{64a_1c_1(d+\sqrt{m}b_2)^2}
{3[(5m-1)b_1 b_2-(8\sqrt{m}-3m-1)d^2- 8db_1(1-\sqrt{m})]}\,.
\eea
For $d=0$ the last constraint reduces to the constraint (\ref{2.9}).

At $m=1$ these solutions go over to the topological (bright-bright) 
solutions (\ref{2.10}) satisfying the constraints (\ref{2.11}) to 
(\ref{2.13}).

{\bf Energy}: Corresponding to the periodic solutions (\ref{v.6}), 
the energy $\hat{E}$ and the constant $C$ are given by
\bea\label{v.14}
&&\hat{E}=\frac{(A^2+F^2)B}{4}E(k)-\frac{BF^2}{4}k'^2K(k)\,, \nonumber \\
&&C=\frac{1}{4}[a_1A^2+a_2F^2]-\frac{B^2}{16}[(3+k^2)A^2+(1+3k^2)F^2]\,.
\eea
On using the expansion formulas (\ref{2.15}) and (\ref{2.16}),
the energy of the periodic solutions can be rewritten 
as the energy of the corresponding hyperbolic (bright-bright) soliton 
solutions [Eq. (\ref{2.10})]  
plus the interaction energy. We find
\be\label{v.17}
\hat{E}=E_{kink}+E_{int}=(A^2+F^2)B \left(\frac{1}{4}
+\frac{k'^2}{8}\left[\ln\left(\frac{4}{k'}\right)-\frac{1}{2} \right] 
\right )-\frac{BF^2}{4}k'^2 \ln \left(\frac{4}{k'}\right)\,. 
\ee
Note that these solutions exist only when $b_1 b_2 > d^2$.  The 
interaction energy vanishes at $k=1$, as it should.

\subsection{Solutions IX and X}

Finally, there are two more mixed type of solutions at $T=T_c$ given by
\be\label{u.6}
\phi=A\sqrt{1\pm\sn(Bx+x_0,m)}\,,~~\psi=F\sqrt{1\mp
\sqrt{m}\sn(Bx+x_0,m)}\,,
\ee
provided the following six coupled equations are satisfied
\be\label{u11}
a_1-b_1 A^2+dF^2+c_1 A^4 = \frac{-B^2}{4}\,,
\ee
\be\label{u12}
-b_1 A^2-\sqrt{m}dF^2+2c_1 A^4 = \frac{-mB^2}{2}\,,
\ee
\be\label{u13}
c_1 A^4 = \frac{3mB^2}{4}\,,
\ee
\be\label{u14}
a_2-b_2 F^2+dA^2+c_2 F^4 = \frac{-mB^2}{4}\,,
\ee
\be\label{u15}
-\sqrt{m}b_2 F^2-dA^2+2\sqrt{m}c_2 F^4 = \frac{-\sqrt{m}B^2}{2}\,,
\ee
\be\label{u16}
c_2 F^4 = \frac{3B^2}{4}\,.
\ee
Three of these equations determine the three unknowns $A,F,B$
while the other three equations give three constraints between the seven
parameters $a_{1,2},b_{1,2},c_{1,2},d$. In particular, $A,F,B$ are given
by
\bea\label{u.7}
&&B^2=\frac{4a_1(b_1 b_2-d^2)}{[(5m-1)b_1 b_2+(11m+1)d^2-
16\sqrt{m}db_1]}\,, \nonumber \\
&&A^2=\frac{8\sqrt{m}(\sqrt{m}b_2-d)a_1}
{[(5m-1)b_1 b_2+(11m+1)d^2- 16\sqrt{m}db_1]}\,, \nonumber \\
&&F^2=\frac{8\sqrt{m}(b_1-\sqrt{m}d)a_1}
{[(5m-1)b_1 b_2+(11m+1)d^2- 16\sqrt{m}db_1]}\,, 
\eea
while the three constraints are
\bea\label{u.8}
&&[(5m-1)b_1 b_2+(11m+1)d^2- 16d\sqrt{m}b_1]a_2= \nonumber \\
&&[(8\sqrt{m}-3-m)b_1 b_2+(8\sqrt{m}+3+m)d^2
-16mdb_2]a_1\,, \nonumber \\
&&c_1(\sqrt{m}b_2-d)^2=mc_2(b_1-\sqrt{m}d)^2\,,~~
(b_1b_2-d^2) =\frac{64a_1c_1(\sqrt{m}b_2-d)^2}
{3[(5m-1)b_1 b_2+(11m+1)d^2- 16d\sqrt{m}b_1]}\,. \nonumber \\
\eea
For $d=0$ the last constraint reduces to the constraint (\ref{2.9}).

At $m=1$ these solutions go over to the
topological (bright-bright) solutions (\ref{y.10})
satisfying the constraints (\ref{y.11}) and (\ref{y.12}).

{\bf Energy}: 
The energy $\hat{E}$ and the constant $C$ for the solutions (\ref{u.6})
are the same as for solutions VII and VIII and are given by Eq. (\ref{v.14})
and hence the interaction energy is again as given by Eq. (\ref{v.17}).
Note that these solutions also exist only when $b_1 b_2 > d^2$.

\subsection{Solution XI}

We shall now present six periodic solutions which exist in the case 
$T < T_c$. Out of these one is bright-bright, three dark-dark and two 
bright-dark solutions. It is easily shown that
\be\label{2.18}
\phi=\frac{A\sn(Bx+x_0,m)}{\sqrt{1-D\sn^2(Bx+x_0,m)}}\,,~~
\psi=\frac{F\sn(Bx+x_0,m)}{\sqrt{1-D\sn^2(Bx+x_0,m)}}\,,
\ee
is a periodic variant of the bright-bright solution provided the 
following six equations are satisfied
\be
a_1=(3D-1-m)B^2\,,
\ee
\be
dF^2-b_1 A^2-2a_1 D=2(m-D-mD)B^2\,,
\ee
\be
-dDF^2+b_1D A^2+a_1 D^2+c_1 A^4=mDB^2\,,
\ee
\be
a_2=(3D-1-m)B^2\,,
\ee
\be
dA^2-b_2 F^2-2a_2 D=2(m-D-mD)B^2\,,
\ee
\be
-dDA^2+b_2D F^2+a_2 D^2+c_2 F^4=mDB^2\,. 
\ee
From here one can determine the four unknowns $A,F,B,D$ and further one
obtains two constraints between the seven parameters
$a_{1,2}.b_{1,2},c_{1,2},d$. We obtain
\bea\label{2.19}
&&B^2=\frac{a_1}{(3D-1-m)}\,,
~~A^2=\frac{2(d+b_2)B^2[2(1+m)D-m-3D^2]}{(b_1b_2-d^2)}\,, \nonumber \\ 
&&F^2=\frac{2(d+b_1)B^2[2(1+m)D-m-3D^2]}{(b_1b_2-d^2)}\,, \nonumber \\ 
&&\frac{[2(1+m)D-m-3D^2]^2}{D(1-D)(m-D)(3D-1-m)}=
\frac{3(b_1b_2-d^2)^2}{4a_1c_1(d+b_2)^2}\,,
\eea
while the two constraints are 
\be\label{2.20}
a_1=a_2\,,~~c_2(d+b_1)^2=c_1(d+b_2)^2\,.
\ee
Using $A^2 >0$ as well as $D < m$, it follows that 
\be
\frac{1+m-\sqrt{1-m+m^2}}{3} < D < m\,.
\ee
As a result, $a_1,a_2$ are positive so long as
\be
\frac{1+m-\sqrt{1-m+m^2}}{3} < D < \frac{1+m}{3}\,,
\ee
while they are negative if 
\be
\frac{1+m}{3} < D < m\,.
\ee

In the limit $d=0$, one obtains the uncoupled kink lattice solution
\cite{behera,falk,sanati} 
\be
\phi=\frac{A\sn(Bx+x_0,m)}{\sqrt{1-D \sn^2(Bx+x_0,m)}}\,,
\ee
satisfying
\bea
&&A^2=\frac{2B^2[2D(1+m)-m-3D^2]}{b_1}\,,
~~B^2=\frac{a_1}{(3D-1-m)}\,, \nonumber \\
&&\frac{[2D(1+m)-m-3D^2]^2}{D(1-D)(m-D)(3D-1-m)}=
\frac{3b_1^2}{4a_1c_1}\,.
\eea

At $m=1$ the solution reduces to the topological (bright-bright) solution 
\be\label{2.21}
\phi=\frac{A\tanh(Bx+x_0)}{\sqrt{1-D\tanh^2(Bx+x_0)}}\,,~~
\psi=\frac{F\tanh(Bx+x_0)}{\sqrt{1-D\tanh^2(Bx+x_0)}}\,,
\ee
provided 
\bea
&&B^2=\frac{a_1}{(3D-2)}\,,
~~A^2=\frac{2(d+b_2)B^2(1-D)(3D-1)}{(b_1b_2-d^2)}\,, \nonumber \\ 
&&F^2=\frac{2(d+b_1)B^2(1-D)(3D-1)}{(b_1b_2-d^2)}\,, \nonumber \\
&&D=\frac{1}{3}\bigg [1+\frac{(b_1b_2-d^2)}
{\sqrt{(b_1b_2-d^2)^2-4a_1c_1(d+b_2)^2}} \bigg ]\,,
\eea
while the two constraints are the same as given by Eq. (\ref{2.20}).
Since $A^2>0$, it then follows that $1/3 < D <1$. Further, 
$a_1,a_2>0$ if $2/3<D<1$ while
 $a_1,a_2<0$ in case $1/3<D<2/3$. 
The constraint $D<1$ implies the inequality 
\be\label{2.22}
(b_1b_2-d^2)^2 > (16/3)a_1c_1(d+b_2)^2\,,
\ee
which for $d=0$ reduces to  
\be
b_1^2>(16/3)a_1c_1\,.
\ee
This corresponds to the situation when there are two degenerate absolute 
minima and a local minimum or maximum at $\phi=0$ depending on if $a_1 >0$ 
or $a_1<0$, respectively \cite{behera,falk,sanati}.

{\bf Special case of $b_1 b_2 =d^2$}

One can show that the solution (\ref{2.18}) exists even in the case 
$b_1 b_2 =d^2$. It turns out that such a solution exists only if 
\be\label{2.23}
b_1 A^4 = b_2 F^4\,,
\ee
and further
\bea\label{2.24}
&&a_1=a_2<0\,,~~b_1 c_2 = b_2 c_1\,,~~B^2=\frac{-a_1}{\sqrt{1-m+m^2}}\,,
\nonumber \\
&&D=\frac{(1+m)-\sqrt{1-m+m^2}}{3}\,,~~c_1 A^4=B^2(1-D)[m(1+D)-2D]\,.
\eea

In the limit $m \rightarrow 1$, relation (\ref{2.24}) takes the simpler
form
\be\label{2.25}
B^2=-a_1\,,~~D=\frac{1}{3}\,,~~c_1 A^4=-\frac{4a_1}{9}\,.
\ee

{\bf Energy}: Corresponding to the periodic solution (\ref{2.18}), 
the energy $\hat{E}$ and the constant $C$ (using the appropriate 
integrals given in \cite{bf}) are given by  
\bea\label{2.26}
&&\hat{E}=(A^2+F^2)B I_1\,, \nonumber \\
&&C=-\frac{1}{2}(A^2+F^2)B^2\,,
\eea
where
\be\label{2.26a}
I_1=\frac{[A_1 K(k)+A_2 E(k)+A_3 \Pi (D,k)]}{4D^2(1-D)(k^2-D)}\,,
\ee
with
\bea\label{2.26b}
&&A_1=(k^2-D)(2D+k^2-3D^2)\,,~~A_2=D[2D(1+k^2)-k^2-3D^2]\,, \nonumber \\
&&A_3 =4(1-D)(k^2-D)[(1+k^2)D-k^2]-[D^2-2(1+k^2)D+3k^2][2(1+k^2)D-k^2-3D^2]\,.
\eea
Here $\Pi(D,k)$ denotes the complete elliptic integral of the third kind 
\cite{gr,bf}.

In order to obtain the interaction energy corresponding to this solution,
one needs the expansion formula for $\Pi(D,k)$ around $k=1$.  Unfortunately, 
only the leading term in the expansion is given in various books that 
we have encountered, but the subleading term is not mentioned.  In 
particular, the result well known in the literature is \cite{bf}
\be\label{kp1}
K(k=1)-(1-\alpha^2)\Pi(\alpha^2,k=1)=\frac{\alpha}{2}
\ln\left[\frac{1+\alpha}{1-\alpha}\right]\,.
\ee
We now show that starting from the basic definitions of $K(k)$ and
$\Pi(\alpha^2,k)$, the subleading term in the expansion is easily derived.

Around $k=1$, we have (note $k'^2=1-k^2$)
\be\label{kp2}
K(k)-(1-\alpha^2)\Pi(\alpha^2,k)=K(1)-(1-\alpha^2)\Pi(\alpha^2,1)+(k^2-1)
\left[\frac{dK}{dk^2}-(1-\alpha^2)\frac{d\Pi}{dk^2}\right]_{k^2=1}+O(k'^4)\,,
\ee
where $K(k)$ and $\Pi(\alpha^2,k)$ are defined by \cite{gr,bf}
\be\label{kp3}
K(k)=\int_{0}^{1} \frac{dx}{\sqrt{(1-x^2)(1-k^2x^2)}}\,,~~
\Pi(\alpha^2,k)=\int_{0}^{1} \frac{dx}{(1-\alpha^2x^2)
\sqrt{(1-x^2)(1-k^2x^2)}}\,.
\ee
The well known result (\ref{kp1}) is easily obtained by using basic
definitions (\ref{kp3}) of $K(k)$ and $\Pi(\alpha^2,k)$, i.e.
\be\label{kp4}
K(1)-(1-\alpha^2)\Pi(\alpha^2,1)= \alpha^2 \int_{0}^{1}
\frac{dx}{(1-\alpha^2x^2)}=\frac{\alpha}{2} \ln\left[\frac{1+\alpha} 
{1-\alpha}\right]\,.
\ee
Similarly, using the basic definitions (\ref{kp3}) it is straightforward 
to show that
\be\label{kp5}
\left[\frac{dK}{dk^2}-(1-\alpha^2)\frac{d\Pi}{dk^2}\right]_{k^2=1}
=\frac{\alpha^2}{2(1-\alpha^2)}\left[\int_{0}^{1}\frac{dx}{(1-x^2)}-
\int_{0}^{1}\frac{dx}{(1-\alpha^2x^2)}\right]\,.
\ee
These integrals are easily evaluated and thus we find that
\bea\label{kp6}
&&K(k)-(1-\alpha^2)\Pi(\alpha^2,k)= 
\frac{\alpha}{2} \ln\left[\frac{1+\alpha}{1-\alpha}\right] \nonumber \\
&&-\frac{k'^2\alpha}{2(1-\alpha^2)}\bigg(\alpha\ln\left[\frac{4}{k'}\right] 
-\frac{1}{2}\ln\left[\frac{1+\alpha}{1-\alpha}\right]\bigg)+O(k'^4)\,.
\eea

On using the expansions of $E(k)$, $K(k)$, and $\Pi(D,k)$ around $k=1$
as given by Eqs. (\ref{2.15}), (\ref{2.16}) and (\ref{kp6}), 
the energy of the periodic solution can be rewritten 
as the energy of the corresponding hyperbolic (bright-bright) soliton 
solution [Eq. (\ref{2.21})]  
plus the interaction energy. We find
\be\label{2.29}
\hat{E}=E_{kink}+E_{int}=(A^2+F^2)B[I_1^{(0)}+k'^2 I_1^{(1)}]\,,
\ee
where $I_1^{(0)}$ and $I_1^{(1)}$ are given by
\be\label{kp7}
I_1^{(0)}=\frac{(3D-1)}{4D(1-D)}+\frac{(1+3D)}{8D^{3/2}}
\ln\left[\frac{1+\sqrt{D}}{1-\sqrt{D}}\right]\,,
\ee
\be\label{kp8}
I_1^{(1)}=\frac{(1+3D^2)}{16D(1-D)^2}-\frac{1}{16D^{3/2}}
\ln\left[\frac{1+\sqrt{D}}{1-\sqrt{D}}\right]\,.
\ee

Note that this solution exists only when $b_1 b_2 \ge d^2$.
The interaction energy, as expected, vanishes at $k=1$.

\subsection{Solution XII}

There are actually three periodic solutions, all of which at $m=1$ reduce 
to the same hyperbolic nontopological (dark-dark) soliton solution. 
However, for $m<1$, all three (periodic) solutions are distinct which we 
now discuss one by one. One of the solution is
\be\label{2.30}
\phi=\frac{A\cn(Bx+x_0,m)}{\sqrt{1-D\sn^2(Bx+x_0,m)}}\,,~~
\psi=\frac{F\cn(Bx+x_0,m)}{\sqrt{1-D\sn^2(Bx+x_0,m)}}\,,
\ee
provided the following six equations are satisfied
\be
a_1-b_1 A^2+dF^2+c_1 A^4=-(1-D)B^2\,,
\ee
\be
(1+D)b_1 A^2-(1+D)d F^2-2a_1 D-2c_1 A^4=2(m-D)(1-D)B^2\,,
\ee
\be
dDF^2-b_1D A^2+a_1 D^2+c_1 A^4=mD(1-D)B^2\,,
\ee
\be
a_2-b_2 F^2+dA^2+c_2 F^4=-(1-D)B^2\,,
\ee
\be
(1+D)b_2 F^2-(1+D)d A^2-2a_2 D-2c_2 F^4=2(m-D)(1-D)B^2\,,
\ee
\be
dDA^2-b_2D F^2+a_2 D^2+c_2 F^4=mD(1-D)B^2\,.
\ee
From here one can determine the four unknowns $A,F,B,D$ and further one
obtains two constraints between the seven parameters
$a_{1,2},b_{1,2},c_{1,2},d$. We get  
\bea\label{2.31}
&&A^2=\frac{2(d+b_2)B^2[(1+2D)m-(D+2)D]}{(1-D)(b_1b_2-d^2)}\,, 
~~F^2=\frac{2(d+b_1)B^2[2(1+2D)m-(D+2)D]}{(1-D)(b_1b_2-d^2)}\,, \nonumber
\\
&&B^2=\frac{(1-D)a_1}{[2m-1-D(2-m)]}\,,
~~\frac{[(1+2D)m-(2+D)D]^2}{D(m-D)[2m-1-D(2-m)]}=
\frac{3(b_1b_2-d^2)^2}{4a_1c_1(d+b_2)^2}\,,
\eea
while the two constraints are 
\be\label{2.32}
a_1=a_2\,,~~c_2(d+b_1)^2=c_1(d+b_2)^2\,.
\ee
Using $A^2 >0$, it follows that 
\be
0 < D <\sqrt{1-m+m^2}-(1-m)\,.
\ee
It is worth noting that for $m < 1/2$, $a_1,a_2 <0$ irrespective of the value
of $D$. 

In the limit $d=0$, one obtains the uncoupled pulse lattice solution
\be
\phi=\frac{A\cn(Bx+x_0,m)}{\sqrt{1-D \sn^2(Bx+x_0,m)}}\,,
\ee
satisfying
\bea
&&A^2=\frac{2B^2[(1+2D)m-(D+2)D]}{(1-D)b_1}\,,
~~B^2=\frac{(1-D)a_1}{[2m-1-D(2-m)]}\,, \nonumber \\
&&\frac{[(1+2D)m-(2+D)D]^2}{D(m-D)[2m-1-D(2-m)]}=
\frac{3b_1^2}{4a_1c_1}\,.
\eea
So far as we are aware of, this solution has never been explicitly 
written down before in the literature.

In the limit $m=1$, the solution (\ref{2.30})
goes over to the hyperbolic, nontopological, dark-dark soliton solution
\be\label{2.33}
\phi=\frac{A\sech(Bx+x_0)}{\sqrt{1-D\tanh^2(Bx+x_0)}}\,,~~
\psi=\frac{F\sech(Bx+x_0)}{\sqrt{1-D\tanh^2(Bx+x_0)}}\,,
\ee
provided 
\bea\label{2.34}
&&B^2=a_1\,,
~~A^2=\frac{2(d+b_2)B^2(1+D)}{(b_1b_2-d^2)}\,, \nonumber \\ 
&&F^2=\frac{2(d+b_1)B^2(1+D)}{(b_1b_2-d^2)}\,, 
\frac{(1+D)^2}{D}=\frac{3(b_1b_2-d^2)^2}
{4a_1c_1(d+b_2)^2}\,,
\eea
while the two constraints are the same as given by Eq. (\ref{2.32}).
Note that $a_1,a_2$ are now positive. The condition $D<1$ gives the
same constraint as given by Eq. (\ref{2.22}). 

{\bf Special case of $b_1 b_2 =d^2$}

One can show that the solution (\ref{2.30}) exists even in the case 
$b_1 b_2 =d^2$. It turns out that such a solution exists only if 
\be\label{2.35}
b_1 A^4 = b_2 F^4\,,
\ee
and further if
\bea\label{2.36}
&&a_1=a_2<0\,,~~b_1 c_2 = b_2 c_1\,,~~~B^2=\frac{-a_1}{\sqrt{1-m+m^2}}\,,
\nonumber \\
&&D=\sqrt{1-m+m^2}-(1-m)\,, ~~~ c_1 A^4=-a_1\left[2-\frac{(2-m)} 
{\sqrt{1-m+m^2}}\right]\,.
\eea

In the limit $m \rightarrow 1$, the constraint (\ref{2.36}) gives
\be\label{2.37}
B^2=-a_1\,,~~D=1\,,~~c_1 A^4=-a_1\,.
\ee
Thus, for the case $b_1 b_2=d^2$ the hyperbolic dark-dark soliton solution
(\ref{2.33}) reduces to a constant solution  
\be\label{2.38}
\phi=A\,,~~\psi=F\,.
\ee

{\bf Energy}: Corresponding to the periodic solution (\ref{2.30}), 
the energy $\hat{E}$ and the constant $C$ are given by
\bea\label{2.39}
&&\hat{E}=(A^2+F^2)BI_{2}\,, \nonumber \\
&&C=-\frac{(1-k^2)(A^2+F^2)B^2}{2(1-D)}\,.
\eea
Here
\be\label{2.40}
I_2=\frac{(1-D)[B_1 K(k)+B_2 E(k)+B_3 \Pi (D,k)]}{4D^2(k^2-D)}\,,
\ee
with
\bea\label{2.41}
&&B_1=\frac{-(k^2-D)[(1-D)^2-(1-k^2)(1-4D)]}{(1-D)}\,,
~~B_2=\frac{D[(1-D^2-(1-k^2)(1+2D)]}{(1-D)}\,, \nonumber \\
&&B_3 =4(k^2-D)[(1+D)k^2-D]-\frac{[D^2-2(1+k^2)D+3k^2][k^2-2(1-k^2)D-D^2]}
{(1-D)}\,.
\eea
On using the expansion formulas for $E(k),K(k)$ and $\Pi(D,k)$ around $k=1$ 
as given above, the energy of the periodic solution can be rewritten 
as the energy of the corresponding hyperbolic (dark-dark) soliton 
solution [Eq. (\ref{2.33})]  
plus the interaction energy. We find
\be\label{2.42}
\hat{E}=E_{kink}+E_{int}=(A^2+F^2)B[I_2^{(0)}+k'^2I_2^{(1)}]\,,
\ee
where $I_2^{(0)}$ and $I_2^{(1)}$ are given by
\be\label{kp9}
I_2^{(0)}=\frac{(1+D)}{4D}-\frac{(1-D)^2}{8D^{3/2}}
\ln\left[\frac{1+\sqrt{D}}{1-\sqrt{D}}\right]\,,
\ee
\be\label{kp10}
I_2^{(1)}=\frac{(D^2-4D-1)}{16D(1-D)}-\frac{(3D^2+6D-1)}{16D^{3/2}(1-D)}
\ln\left[\frac{1+\sqrt{D}}{1-\sqrt{D}}\right]+\frac{1}{(1-D)}\ln\left(\frac{4} 
{k'}\right)\,.
\ee

Note that this solution exists only when $b_1 b_2 \ge d^2$.  
The interaction energy, as expected, vanishes at $k=1$.

\subsection{Solution XIII}

Another solution is given by
\be\label{2.43}
\phi=\frac{A\dn(Bx+x_0,m)}{\sqrt{1-D\sn^2(Bx+x_0,m)}}\,,~~
\psi=\frac{F\dn(Bx+x_0,m)}{\sqrt{1-D\sn^2(Bx+x_0,m)}}\,,
\ee
provided the following six equations are satisfied 
\be
a_1-b_1 A^2+dF^2+c_1 A^4=-(m-D)B^2\,,
\ee
\be
(m+D)b_1 A^2-(m+D)d F^2-2a_1 D-2mc_1 A^4=2(m-D)(1-D)B^2\,,
\ee
\be
dmDF^2-b_1mD A^2+a_1 D^2+c_1m^2 A^4=D(m-D)B^2\,,
\ee
\be
a_2-b_2 F^2+dA^2+c_2 F^4=-(m-D)B^2\,,
\ee
\be
(m+D)b_2 F^2-(m+D)d A^2-2a_2 D-2mc_2 F^4=2(m-D)(1-D)B^2\,,
\ee
\be
dmDA^2-b_2mD F^2+a_2 D^2+c_2m^2 F^4=D(m-D)B^2\,.
\ee
From here one can determine the four unknowns $A,F,B,D$ and further one
obtains two constraints between the seven parameters
$a_{1,2},b_{1,2},c_{1,2},d$. We obtain
\bea\label{2.44}
&&A^2=\frac{2(d+b_2)B^2[2D(1-m)+m-D^2]}{(m-D)(b_1b_2-d^2)}\,,~~ 
F^2=\frac{2(d+b_1)B^2[2D(1-m)+m-D^2]}{(m-D)(b_1b_2-d^2)}\,, \nonumber \\  
&&B^2=\frac{(m-D)a_1}{[2m(1-D)-m^2+D]}\,,
~~\frac{[2D(1-m)+m-D^2]^2}{D(1-D)[2m(1-D)-m^2+D]}=
\frac{3(b_1b_2-d^2)^2}{4a_1c_1(d+b_2)^2}\,,
\eea
while the two constraints are 
\be\label{2.45}
a_1=a_2\,,~~c_2(d+b_1)^2=c_1(d+b_2)^2\,.
\ee
The best bound on $D$ is given by $0<D<m$ which ensures $A^2 >0$.
Further, unlike two of the above solutions (i.e. solutions XI and XII),
$a_1,a_2$ are always positive.

In the limit $d=0$, one obtains the uncoupled pulse lattice solution
\be
\phi=\frac{A\dn(Bx+x_0,m)}{\sqrt{1-D \sn^2(Bx+x_0,m)}}\,,
\ee
satisfying
\bea
&&A^2=\frac{2B^2[2D(1-m)+m-D^2]}{(m-D)b_1}\,,
~~B^2=\frac{(m-D)a_1}{[2m(1-D)+D-m^2]}\,, \nonumber \\
&&\frac{[2D(1-m)+m-D^2]^2}{D(1-D)[2m(1-D)+D-m^2]}=
\frac{3b_1^2}{4a_1c_1}\,.
\eea

In the limit $m=1$, the solution (\ref{2.43}) also 
goes over to the hyperbolic, nontopological, dark-dark soliton solution
(\ref{2.33}) satisfying the same constraint as given by Eq.
(\ref{2.34}).

{\bf Special case of $b_1 b_2 =d^2$}

One can show that the solution (\ref{2.43}) exists even in the case 
$b_1 b_2 =d^2$. It turns out that such a solution exists only if 
\be\label{2.46}
b_1 A^4 = b_2 F^4\,,
\ee
and further if
\bea\label{2.47}
&&a_1=a_2\,,~~b_1 c_2 = b_2 c_1\,,~~~B^2=\frac{-a_1}{\sqrt{1-m+m^2}}\,,
\nonumber \\
&&D=\sqrt{1-m+m^2}+(1-m)\,, ~~~c_1 A^4=\frac{3D(1-D)B^2}{(m-D)}\,.
\eea
In the limit $m \rightarrow 1$, the constraint (\ref{2.47}), as expected,
reduces to Eq. (\ref{2.37}) and the solution (\ref{2.43}) reduces to the
constant solution (\ref{2.38}).

{\bf Energy}: Corresponding to the periodic solution (\ref{2.43}), 
the energy $\hat{E}$ and the constant $C$ are given by
\bea\label{2.48}
&&\hat{E}=(A^2+F^2)BI_{3}\,, \nonumber \\
&&C=\frac{k^2(1-k^2)(A^2+F^2)B^2}{2(k^2-D)}\,.
\eea
Here
\be\label{2.49}
I_3=\frac{(k^2-D)[D_1 K(k)+D_2 E(k)+D_3 \Pi (D,k)]}{4D^2(1-D)}\,,
\ee
with
\bea\label{2.50}
&&D_1=-[(1-D)^2-(1-k^2)]\,,
~~D_2=\frac{D[(1-D^2+(1-k^2)(2D-1)]}{(k^2-D)}\,, \nonumber \\
&&D_3 =4(1-D)[(1-D)k^2+D]-\frac{[D^2-2(1+k^2)D+3k^2][k^2+2(1-k^2)D-D^2]}{(k^2-D)}\,.
\eea
On using the expansion formulas for $E(k),K(k)$ and $\Pi(D,k)$ around 
$k=1$ obtained above, the energy of the periodic solution can be rewritten 
as the energy of the corresponding hyperbolic (dark-dark) soliton 
solution [Eq. (\ref{2.33})]  
plus the interaction energy. We find
\be\label{2.51}
\hat{E}=E_{kink}+E_{int}=(A^2+F^2)B[I_3^{(0)}+k'^2 I_3^{(1)}]\,,
\ee
where $I_3^{(0)}$ is in fact the {\it same} as $I_2^{(0)}$, and given by 
Eq. (\ref{kp9}) while $I_3^{(1)}$ is given by
\be\label{kp11}
I_3^{(1)}=\frac{(D^2+8D-5)}{16D(1-D)}+\frac{(7D^2-2D+3)}{16D^{3/2}(1-D)}
\ln\left[\frac{1+\sqrt{D}}{1-\sqrt{D}}\right]-\frac{1}{(1-D)}\ln\left(\frac{4} 
{k'}\right)\,.
\ee

Note that this solution exists only when $b_1 b_2 \ge d^2$.  
As expected, the interaction energy vanishes at $k=1$.

\subsection{Solution XIV}

Yet another solution is
\be\label{2.52}
\phi=\frac{A\cn(Bx+x_0,m)}{\sqrt{1-D\sn^2(Bx+x_0,m)}}\,,~~
\psi=\frac{F\dn(Bx+x_0,m)}{\sqrt{1-D\sn^2(Bx+x_0,m)}}\,,
\ee
provided the following six equations are satisfied
\be
a_1-b_1 A^2+dF^2+c_1 A^4=-(1-D)B^2\,,
\ee
\be
(1+D)b_1 A^2-(m+D)d F^2-2a_1 D-2c_1 A^4=2(m-D)(1-D)B^2\,,
\ee
\be
dmDF^2-b_1D A^2+a_1 D^2+c_1 A^4=mD(1-D)B^2\,,
\ee
\be
a_2-b_2 F^2+dA^2+c_2 F^4=-(m-D)B^2\,,
\ee
\be
(m+D)b_2 F^2-(1+D)d A^2-2a_2 D-2mc_2 F^4=2(m-D)(1-D)B^2\,,
\ee
\be
dDA^2-b_2mD F^2+a_2 D^2+c_2m^2 F^4=D(m-D)B^2\,.
\ee
Solving these equations, we determine the four unknowns $A,F,B,D$
\bea\label{2.53}
&&B^2=\frac{3D(1-D)(m-D)(b_1b_2-d^2)^2}
{4c_1[(m-D^2)(d+b_2)-2D(1-m)(b_2-d)]^2}\,, \nonumber \\
&&A^2=\frac{2B^2[(m-D^2)(d+b_2)-2D(1-m)(b_2-d)]}{(b_1b_2-d^2)(1-D)}\,, 
\nonumber
\\
&&F^2=\frac{2B^2[(m-D^2)(d+b_1)+2D(1-m)(b_1-d)]}{(b_1b_2-d^2)(m-D)}\,,
\nonumber \\
&&(1-D)a_1+(1-m)dF^2=[2m-1-(2-m)D]B^2\,,
\eea
while the two constraints between the seven parameters are
\be\label{2.54}
(1-D)^2 c_1 A^4= (m-D)^2 c_2 F^4\,,
\ee
and
\be\label{2.55}
(1-D)a_2=(1-m)b_2F^2+\frac{[D^2+4mD-3D(1+m^2)+m^2]B^2}{(m-D)}\,.  
\ee
In this case too the best bound on $D$ is given by $0<D<m$.
Further, $a_1 <0$ in case $m <1/2$.

In the limit $m=1$, the solution (\ref{2.52}) also 
goes over to the hyperbolic, nontopological, dark-dark soliton solution
(\ref{2.33}) satisfying the same constraints as given by Eq. (\ref{2.34}).

Unlike the solutions XI, XII and XIII, the solution (\ref{2.52}) does not 
exist in the case $b_1 b_2 =d^2$. This is because, in order that such a 
solution exist, $m$ and $D$ must satisfy
\be
\sqrt{b_2}[m-D^2-2(1-m)D]=-\sqrt{b_1}[m-D^2+2(1-m)D]\,,
\ee
which is impossible.

{\bf Energy}: Corresponding to the periodic solution (\ref{2.52}), 
the energy $\hat{E}$ and the constant $C$ are given by
\bea\label{2.58}
&&\hat{E}=B(A^2I_2+F^2I_{3})\,, \nonumber \\
&&C=-\frac{1}{2D^3}\bigg [A^2(-a_1D^2+\frac{b_1}{2}DA^2-\frac{c_1}{3}A^4
-dmDF^2) 
+F^2(-ma_2D^2+\frac{b_2}{2}Dm^2F^2-\frac{c_2}{3}m^3F^4) \bigg ]\,,
\eea
where $I_2,I_3$ are as given by Eqs. (\ref{2.40}), (\ref{2.41}),
(\ref{2.49}) and (\ref{2.50}). 
On using the expansion formulas for $E(k),K(k)$ and $\Pi(D,k)$ around 
$k=1$ as derived above, the energy of the periodic solution can be rewritten 
as the energy of the corresponding hyperbolic (dark-dark) soliton 
solution [Eq. (\ref{2.33})]  
plus the interaction energy. We find
\be\label{2.59}
\hat{E}=E_{kink}+E_{int}=[(A^2+F^2)I_2^{(0)}+k'^2(A^2I_2^{(1)} 
+F^2I_3^{(1)})]B\,,
\ee
where $I_2^{(0)},I_2^{(1)},I_3^{(1)}$ are given by Eqs. (\ref{kp9}),  
(\ref{kp10}) and (\ref{kp11}) respectively.

Note that this solution exists only when $b_1 b_2 > d^2$.  
As expected, the interaction energy vanishes at $k=1$.

\subsection{Solution XV}

Finally, we obtain two periodic solutions which at $m=1$
go over to the same bright-dark soliton solution. The first 
solution is given by
\be\label{2.60}
\phi=\frac{A\sn(Bx+x_0,m)}{\sqrt{1-D\sn^2(Bx+x_0,m)}}\,,~~
\psi=\frac{F\cn(Bx+x_0,m)}{\sqrt{1-D\sn^2(Bx+x_0,m)}}\,,~~
\ee
provided the following six equations are satisfied
\be
a_1+dF^2=(3D-1-m)B^2\,,
\ee
\be
b_1 A^2+d(1+D)F^2+2a_1 D=2(D+Dm-m)B^2\,,
\ee
\be
dDF^2+b_1D A^2+a_1 D^2+c_1 A^4=mDB^2\,,
\ee
\be
a_2-b_2 F^2+c_2 F^4=-(1-D)B^2\,,
\ee
\be
(1+D)b_2 F^2+d A^2-2a_2 D-2c_2 F^4=2(m-D)(1-D)B^2\,,
\ee
\be
-dDA^2-b_2D F^2+a_2 D^2+c_2 F^4=mD(1-D)B^2\,.
\ee
On solving these equations we determine the four parameters $A,F,B,D$
as well as obtain two constraints between the seven parameters. We find
\bea\label{2.61a}
&&A^2=\frac{2B^2[b_2(2mD-m+2D-3D^2)-d(2mD+m-2D-D^2)]}{(b_1b_2-d^2)}\,, 
\nonumber
\\
&&F^2=\frac{2B^2[b_1(2mD+m-2D-D^2)-d(2mD-m+2D-3D^2)]}{(1-D)(b_1b_2-d^2)}\,,
\eea
\bea\label{2.61}
&&B^2=\frac{3D(1-D)(m-D)(b_1b_2-d^2)^2}
{4c_1[b_2(2mD-m+2D-3D^2)-d(2mD+m-2D-D^2)]^2}\,, \nonumber \\
&&a_1+dF^2=[3D-1-m]B^2\,,
\eea
while the two constraints are
\be\label{2.62}
c_1 A^4=(1-D)^2 c_2 F^4\,,
\ee
\be\label{2.63}
a_2-b_2F^2+\frac{B^2[(1-D)^2+3D(m-D)]}{(1-D)}=0\,.
\ee

From positivity considerations, one can show from here that
\be
\frac{(1+m)-\sqrt{1-m+m^2}}{3} < D < \sqrt{1-m+m^2}-(1-m)\,.
\ee
Further, it follows that $a_1 < 0$ if $D<(1+m)/3$.

At m=1, the solution (\ref{2.60}) goes over to the
hyperbolic, bright-dark soliton solution
\be\label{2.64}
\phi=\frac{A\tanh(Bx+x_0)}{\sqrt{1-D\tanh^2(Bx+x_0)}}\,,~~
\psi=\frac{F\sech(Bx+x_0)}{\sqrt{1-D\tanh^2(Bx+x_0)}}\,.\,
\ee
and the relations (\ref{2.61a}) to (\ref{2.63}) take the simpler form
\bea\label{2.65a}
&&A^2=\frac{2B^2(1-D)[b_2(3D-1)-d(1+D)]}{(b_1b_2-d^2)}\,, \nonumber
\\
&&F^2=\frac{2B^2(1-D)[b_1(1+D)-d(3D-1)]}{(b_1b_2-d^2)}\,,
\eea
\bea\label{2.65}
&&B^2=\frac{3D(b_1b_2-d^2)^2}
{4c_1[b_2(3D-1)-d(1+D)]^2}\,, \nonumber \\
&&a_1+dF^2=(3D-2)B^2\,,
\eea
while the two constraints are
\be\label{2.66}
c_1 A^4=(1-D)^2 c_2 F^4\,,
\ee
\be\label{2.67}
a_2-b_2F^2+B^2(1+2D)=0\,.
\ee
From here one can show that $1/3 < D < 1$ and further $a_1 <0$ if
$D>2/3$ while $a_2 <0$ if $D > \frac{\sqrt{3}-1}{2}$.

{\bf Special case of $b_1 b_2 =d^2$}

One can show that the solution (\ref{2.60}) exists even in the case 
$b_1 b_2 =d^2$. It turns out that such a solution exists only if
\be\label{2.68}
\sqrt{b_2}[2(1+m)D-m-3D^2]=\sqrt{b_1}[m-D^2-2(1-m)D]\,,
\ee
and further while relations (\ref{2.61}) to  (\ref{2.63})
are still valid, the relations (\ref{2.61a}) are no longer valid. Instead
of separate expressions for $A^2,F^2$, one now only has a constraint
\be\label{2.69}
b_1A^2+\sqrt{b_1 b_2}(1-D)F^2=2B^2[2(1+m)D-3D^2-m]\,.
\ee

In the limit $m \rightarrow 1$, the hyperbolic bright-dark solution
(\ref{2.64}) is still valid provided $D$ has the value 
\be\label{2.70}
D=\frac{\sqrt{b_2}+\sqrt{b_1}}{3\sqrt{b_2}-\sqrt{b_1}}\,,
\ee
while the constraint (\ref{2.69}) takes the simpler form
\be\label{2.71}
b_1A^2+\sqrt{b_1 b_2}(1-D)F^2=2B^2(1-D)(3D-1)\,.
\ee
The relations (\ref{2.65}) to (\ref{2.67}) are still valid.

{\bf Energy}: Corresponding to the periodic solution (\ref{2.60}), 
the energy $\hat{E}$ and the constant $C$ are given by
\bea\label{2.72}
&&\hat{E}=B(A^2I_1+F^2I_{2})\,, \nonumber \\
&&C=-\frac{1}{2D^3}\bigg [A^2(a_1D^2+\frac{b_1}{2}DA^2+\frac{c_1}{3}A^4
+dDF^2) 
+F^2(-a_2D^2+\frac{b_2}{2}DF^2-\frac{c_2}{3}F^4) \bigg ]\,,
\eea
where $I_1,I_2$ are as given by Eqs. (\ref{2.26a}), (\ref{2.26b}),
(\ref{2.40}) and (\ref{2.41}). 
On using the expansion formulas for $E(k),K(k)$ and $\Pi(D,k)$ around 
$k=1$ as derived above, the energy of the periodic solution can be rewritten 
as the energy of the corresponding hyperbolic (bright-dark) soliton 
solution [Eq. (\ref{2.64})]  
plus the interaction energy. We find
\be\label{2.73}
\hat{E}=E_{kink}+E_{int}=BA^2[I_1^{(0)}+k'^2I_1^{(1)}]
+BF^2[I_2^{(0)}+k'^2I_2^{(1)}]\,,
\ee
where $I_{1,2}^{(0,1)}$ are given by Eqs. (\ref{kp7}), (\ref{kp8}),
(\ref{kp9}) and (\ref{kp10}).

Note that this solution exists only when $b_1 b_2 \ge d^2$.  As 
usual, the interaction energy vanishes at $k=1$.

\subsection{Solution XVI}

Finally, the last solution below $T_c$ is given by
\be\label{2.74}
\phi=\frac{A\sn(Bx+x_0,m)}{\sqrt{1-D\sn^2(Bx+x_0,m)}}\,,~~
\psi=\frac{F\dn(Bx+x_0,m)}{\sqrt{1-D\sn^2(Bx+x_0,m)}}\,,~~
\ee
provided the following six equations are satisfied
\be
a_1+dF^2=(3D-1-m)B^2\,,
\ee
\be
b_1 A^2+d(m+D)F^2+2a_1 D=2(D+Dm-m)B^2\,,
\ee
\be
dmDF^2+b_1D A^2+a_1 D^2+c_1 A^4=mDB^2\,,
\ee
\be
a_2-b_2 F^2+c_2 F^4=-(m-D)B^2\,,
\ee
\be
(m+D)b_2 F^2+d A^2-2a_2 D-2mc_2 F^4=2(m-D)(1-D)B^2\,,
\ee
\be
-dDA^2-mb_2D F^2+a_2 D^2+m^2 c_2 F^4=D(1-D)B^2\,.
\ee
On solving these equations we obtain the four unknowns $A,F,B,D$ and
further also obtain two constraints between the seven parameters. We find
\bea\label{2.75}
&&A^2=\frac{2B^2[b_2(2mD-m+2D-3D^2)-d(m+2D-2mD-D^2)]}{(b_1b_2-d^2)}\,, 
\nonumber
\\
&&F^2=\frac{2B^2[b_1(2D+m-2Dm-D^2)-d(2mD-m+2D-3D^2)]}{(m-D)(b_1b_2-d^2)}\,,
\eea
\be\label{2.76}
B^2=\frac{3D(1-D)(m-D)}{4c_1[b_2(2mD-m+2D-3D^2)-d(m+2D-2mD-D^2)]^2}\,,
~~a_1+dF^2=[3D-1-m]B^2\,,
\ee
while the two constraints are
\be\label{2.77}
c_1 A^4=(m-D)^2 c_2 F^4\,,
\ee
\be\label{2.78}
a_2-b_2F^2+\frac{B^2[(m-D)^2+3D(1-D)]}{(m-D)}=0\,.
\ee
From positivity considerations, one can show from here that
\be
\frac{(1+m)-\sqrt{(1+m)(m-1/2)}}{3} < D < m\,.
\ee
Further, it follows that $a_1 < 0$ if $D<(1+m)/3$.

At m=1, the solution (\ref{2.74}) goes over to the  
hyperbolic, bright-dark soliton solution (\ref{2.64})
and the constraints (\ref{2.75}) to (\ref{2.78}) take the simpler form
as given by Eqs. (\ref{2.65a}) to (\ref{2.67}).

{\bf Special case of $b_1 b_2 =d^2$}

One can show that the solution (\ref{2.74}) exists even in the case 
$b_1 b_2 =d^2$. It turns out that such a solution exists only if
\be\label{2.79}
\sqrt{b_2}[2(1+m)D-m-3D^2]=\sqrt{b_1}[m-D^2+2(1-m)D]\,,
\ee
and further while relations (\ref{2.76}) to  (\ref{2.78})
are still valid, the relation (\ref{2.75}) is no longer valid. Instead
of separate expressions for $A^2,F^2$, one now only has a constraint
\be\label{2.80}
b_1A^2+\sqrt{b_1 b_2}(1-D)F^2=2B^2[2(1+m)D-3D^2-m]\,.
\ee

In the limit $m \rightarrow 1$, the hyperbolic bright-dark solution
(\ref{2.64}) is still valid provided the constraints (\ref{2.70}) and
(\ref{2.71}) as well as Eqs. (\ref{2.65}) to
(\ref{2.67}) are satisfied.

{\bf Energy}: Corresponding to the periodic solution (\ref{2.74}), 
the energy $\hat{E}$ and the constant $C$ are given by
\bea\label{2.81}
&&\hat{E}=B(A^2I_1+F^2I_{3})\,, \nonumber \\ 
&&C=-\frac{1}{2D^3}\bigg [A^2(a_1D^2+\frac{b_1}{2}DA^2+\frac{c_1}{3}A^4
+dmDF^2) 
+F^2(-ma_2D^2+\frac{b_2}{2}Dm^2F^2-\frac{c_2}{3}m^3F^4) \bigg ]\,,
\eea
where $I_1,I_3$ are as given by Eqs. (\ref{2.26a}), (\ref{2.26b}),
(\ref{2.49}) and (\ref{2.50}). 
On using the expansion formulas for $E(k),K(k)$ and $\Pi(D,k)$ around 
$k=1$ as derived above, the energy of the periodic solution can be rewritten 
as the energy of the corresponding hyperbolic (bright-dark) soliton 
solution [Eq. (\ref{2.64})]  
plus the interaction energy. We find
\be\label{2.82}
\hat{E}=E_{kink}+E_{int}=BA^2[I_1^{(0)}+k'^2I_1^{(1)}]
+BF^2[I_3^{(0)}+k'^2I_3^{(1)}]\,,
\ee
where $I_1^{(0,1)}$ are given by Eqs. (\ref{kp7}) and (\ref{kp8}) while
$I_3^{(0)}=I_2^{(0)}$ and $I_3^{(1)}$ are  given by Eqs. (\ref{kp9})
and (\ref{kp11}), respectively.

Note that this solution exists only when $b_1 b_2 \ge d^2$.  The 
interaction energy vanishes at $k=1$.  

\subsection{Solution XVII}

So far, we have presented ten solutions at $T=T_c$ and six solutions for  
$T < T_C$. We now present one solution which exists when $T_c < T < T_p$
where $T_p$ denotes the point of inflection where $\phi^6$ potential (at
$d=0$) has an absolute minimum at $\phi=0$ and two points of inflection. 
Note that the point of inflection occurs when $b^2 = 4a_1c_1$.

It is easily shown that
\be\label{u.18}
\phi=\frac{A}{\sqrt{1-D\sn^2(Bx+x_0,m)}}\,,~~
\psi=\frac{F}{\sqrt{1-D\sn^2(Bx+x_0,m)}}\,,
\ee
is a solution to the field Eqs. (\ref{2.2}) 
provided the following six equations are satisfied
\be
a_1-b_1A^2+c_1A^4+dF^2=DB^2\,,
\ee
\be
-dF^2+b_1 A^2-2a_1 =2(D-1-m)B^2\,,
\ee
\be
a_1 D=(3m-D-Dm)B^2\,,
\ee
\be
a_2-b_2F^2+c_2F^4+dA^2=DB^2\,,
\ee
\be
-dA^2+b_2 F^2-2a_2 =2(D-1-m)B^2\,,
\ee
\be
a_2 D=(3m-D-Dm)B^2\,.  
\ee
From here one can determine the four unknowns $A,F,B,D$ and further one
obtains two constraints between the seven parameters
$a_{1,2},b_{1,2},c_{1,2},d$. We get 
\bea\label{u.19}
&&B^2=\frac{a_1D}{[3m-D(1+m)]}\,,
~~A^2=\frac{2(d+b_2)B^2[D^2-2(1+m)D+3m]}{D(b_1b_2-d^2)}\,, \nonumber \\ 
&&F^2=\frac{2(d+b_1)B^2[D^2-2(1+m)D+3m]}{D(b_1b_2-d^2)}\,, \nonumber \\ 
&&\frac{[D^2-2(1+m)D+3m]^2}{(1-D)(m-D)[3m-D(1+m)]}=
\frac{3(b_1b_2-d^2)^2}{4a_1c_1(d+b_2)^2}\,,
\eea
while the two constraints are 
\be\label{u.20}
a_1=a_2\,,~~c_2(d+b_1)^2=c_1(d+b_2)^2\,.
\ee
Note that $a_1,a_2$ are always positive.

In the limit $d=0$, one obtains the uncoupled pulse lattice solution
\cite{behera,falk,sanati} 
\be
\phi=\frac{A}{\sqrt{1-D \sn^2(Bx+x_0,m)}}\,,
\ee
satisfying
\bea
&&A^2=\frac{2B^2[D^2-2D(1+m)+3m]}{Db_1}\,,
~~B^2=\frac{(1-D)a_1}{[2m-1-D(2-m)]}\,, \nonumber \\
&&\frac{[D^2-2D(1+m)+3m]^2}{(1-D)(m-D)[3m-(1+m)D]}=
\frac{3b_1^2}{4a_1c_1}\,.
\eea

At $m=1$ the solution reduces to the nontopological (dark-dark) solution
\be\label{u.21}
\phi=\frac{A}{\sqrt{1-D\tanh^2(Bx+x_0)}}\,,~~
\psi=\frac{F}{\sqrt{1-D\tanh^2(Bx+x_0)}}\,,
\ee
provided 
\bea
&&B^2=\frac{a_1D}{(3-2D)}\,,
~~A^2=\frac{2(d+b_2)B^2(1-D)(3-D)}{D(b_1b_2-d^2)}\,, \nonumber \\ 
&&F^2=\frac{2(d+b_1)B^2(1-D)(3-D)}{D(b_1b_2-d^2)}\,,~~
\frac{(3-D)^2}{(3-2D)}=\frac{3(b_1b_2-d^2)^2}{4a_1c_1(d+b_2)^2}\,,
\eea
while the two constraints are the same as given by Eq. (\ref{u.20}).
The constraint $0<D<1$ implies the relation  
\be\label{u.22}
4a_1c_1(d+b_2)^2 < (b_1b_2-d^2)^2 < (16/3)a_1c_1(d+b_2)^2\,,
\ee
which for $d=0$ reduces to the constraint 
\be
4a_1c_1<\,b_1^2\, <(16/3)a_1c_1\,,
\ee
i.e. $T_c < T < T_p$.
This corresponds to the situation when there is an absolute minimum at 
$\phi=0$ and there are two degenerate local minima \cite{behera,falk,sanati}.

{\bf Special case of $b_1 b_2 =d^2$}

One can show that the solution (\ref{u.18}) exists even in the case 
$b_1 b_2 =d^2$ provided $m <1$. 
It turns out that such a solution exists only if 
\be\label{u.23}
b_1 A^4 = b_2 F^4\,,~~c_2b_1=c_1b_2\,,
\ee
and further
\bea\label{u.24}
D=(1+m)-\sqrt{1-m+m^2}\,,~~c_1 A^4=\frac{3B^2(1-D)(m-D)}{D}\,,~~
a_1=a_2=\frac{B^2[3m-D(1+m)]}{D}\,.
\eea

{\bf Energy}: Corresponding to the periodic solution (\ref{u.18}), 
the energy $\hat{E}$ and the constant $C$ are given by (using appropriate 
integrals in \cite{bf}) 
\bea\label{u.26}
&&\hat{E}=(A^2+F^2)B I_4\,, \nonumber \\
&&C=\frac{m}{2}(A^2+F^2)B^2\,,
\eea
where
\be\label{u.26a}
I_4=\frac{[G_1 K(k)+G_2 E(k)+G_3 \Pi (D,k)]}{4D^2(1-D)(k^2-D)}\,,
\ee
with
\bea\label{u.26b}
&&G_1=D(k^2-D)(D^2-2D+4Dk^2-3k^2)\,,~~G_2=D^2[D^2-2D(1+k^2)+3k^2]\,, \nonumber \\
&&G_3 =4D(1-D)(k^2-D)[3k^2-(1+k^2)D]-D[D^2-2(1+k^2)D+3k^2]^2\,.
\eea

On using the expansion formulas for $E(k),K(k)$ and $\Pi(D,k)$ around
$k=1$ derived above, the energy of the periodic solution can be rewritten 
as the energy of the corresponding hyperbolic (dark-dark) soliton 
solution [Eq. (\ref{u.21})]  
plus the interaction energy. We find
\be\label{u.29a}
\hat{E}=E_{kink}+E_{int}=(A^2+F^2)B[I_4^{(0)}+k'^2 I_4^{(1)}]\,,
\ee
where $I_4^{(0)}$ and $I_4^{(1)}$ are given by
\be\label{kp12}
I_4^{(0)}=\frac{(3-D)}{4(1-D)}-\frac{(D+3)}{8D^{/2}}
\ln\left[\frac{1+\sqrt{D}}{1-\sqrt{D}}\right]\,,
\ee
\be\label{kp13}
I_4^{(1)}=\frac{-(D^2-8D+3)}{16(1-D)^2}+\frac{3(D+4)}{16D^{3/2}}
\ln\left[\frac{1+\sqrt{D}}{1-\sqrt{D}}\right]-\frac{3}{2D^2}\ln\left(\frac{4} 
{k'}\right)\,.
\ee

Note that this solution exists only when $b_1 b_2 \ge d^2$.  The 
interaction energy vanishes at $k=1$.  

\subsection{Solution XVIII}

Finally, we present three novel mixed phase solutions, i.e. 
in which one of the field 
exists for $T> T_c$ while the other one exists for $T<T_c$.

It is easily shown that
\be\label{v.18}
\phi=\frac{A}{\sqrt{1-D\sn^2(Bx+x_0,m)}}\,,~~
\psi=\frac{F\sn(Bx+x_0,m)}{\sqrt{1-D\sn^2(Bx+x_0,m)}}\,,
\ee
is a solution to the field Eqs. (\ref{2.2}) 
provided the following six equations are satisfied
\be
a_1-b_1A^2+c_1A^4=DB^2\,,
\ee
\be
dF^2+b_1 DA^2-2a_1 D=2D(D-1-m)B^2\,,
\ee
\be
a_1 D-dF^2=(3m-D-Dm)B^2\,,
\ee
\be
a_2+dA^2=(3D-1-m)B^2\,,
\ee
\be
-dDA^2-b_2 F^2-2a_2 D=2(m-D-mD)B^2\,,
\ee
\be
a_2 D^2+b_2DF^2+c_2F^4=mDB^2\,.  
\ee
On solving these equations, we obtain the four unknowns $A,F,B,D$ as well
as two constraints between the seven parameters. We find
\bea\label{v.19}
&&A^2=\frac{2B^2[b_2(D^2-2D-2mD+3m)+d(2D+2mD-m-3D^2)]}
{(b_1b_2-d^2)}\,, \nonumber \\
&&F^2=\frac{2B^2[d(D^2-2D-2mD+3m)+b_1(2D+2mD-m-3D^2)]}
{(b_1b_2-d^2)}\,, \nonumber \\
&&B^2=\frac{3(1-D)(m-D)(b_1b_2-d^2)^2}
{4Dc_1[b_2(D^2-2D-2mD+3m)+d(2D+2mD-m-3D^2)]^2}\,, \nonumber \\
&&a_1D-dF^2=[3m-(1+m)D]B^2\,,
\eea
while the two constraints are
\be\label{v.19aa}
D^2c_1A^4=c_2F^4\,,~~a_2+dA^2=(3D-1-m)B^2\,.
\ee

At $m=1$ the solution reduces to the dark-bright solution
\be\label{v.21}
\phi=\frac{A}{\sqrt{1-D\tanh^2(Bx+x_0)}}\,,~~
\psi=\frac{F\tanh(Bx+x_0)}{\sqrt{1-D\tanh^2(Bx+x_0)}}\,,
\ee
provided
\bea
&&A^2=\frac{2B^2(1-D)[b_2(3-D)+d(3D-1)]}
{(b_1b_2-d^2)}\,, \nonumber \\
&&F^2=\frac{2B^2(1-D)[d(3-D)+b_1(3D-1)]}
{(b_1b_2-d^2)}\,, \nonumber \\
&&B^2=\frac{3(b_1b_2-d^2)^2}
{4Dc_1[b_2(3-D)+d(3D-1)]^2}\,,
~~a_1D-dF^2=[3-2D]B^2\,,
\eea
while the two constraints are
\be
D^2c_1A^4=c_2F^4\,,~~a_2+dA^2=(3D-2)B^2\,.
\ee

{\bf Energy}: Corresponding to the periodic solution (\ref{v.18}), 
the energy $\hat{E}$ and the constant $C$ are given by (using 
appropriate integrals in \cite{bf}) 
\bea\label{v.26}
&&\hat{E}=(A^2I_4+F^2I_1)B\,, \nonumber \\
&&C=\frac{m}{2}A^2B^2-\frac{F^2}{4D}[B^2(1+m-D)+a_2]\,,
\eea
where $I_1$ and $I_4$ are given by Eqs. (\ref{2.26a}), (\ref{2.26b}),
(\ref{u.26a}) and (\ref{u.26b}). 

On using the expansion formulas for $E(k),K(k)$ and $\Pi(D,k)$ around
$k=1$ as derived above,
the energy of the periodic solution can be rewritten 
as the energy of the corresponding hyperbolic (dark-bright) soliton 
solution [Eq. (\ref{v.21})]  
plus the interaction energy. We find
\be\label{v.29}
\hat{E}=E_{kink}+E_{int}=BA^2[I_4^{(0)}+k'^2I_4^{(1)}]
+BF^2[I_1^{(0)}+k'^2I_1^{(1)}]\,,
\ee
where $I_{1,4}^{(0,1)}$ are given by Eqs. (\ref{kp7}), (\ref{kp8}),
(\ref{kp12}) and (\ref{kp13}).

Note that this solution exists only when $b_1 b_2 > d^2$.  The interaction 
energy vanishes at $k=1$.

\subsection{Solution XIX}

Another novel mixed phase solution is given by
\be\label{w.18}
\phi=\frac{A}{\sqrt{1-D\sn^2(Bx+x_0,m)}}\,,~~
\psi=\frac{F\cn(Bx+x_0,m)}{\sqrt{1-D\sn^2(Bx+x_0,m)}}\,.
\ee
This is a solution to the field Eqs. (\ref{2.2}) 
provided the following six equations are satisfied
\be
a_1-b_1A^2+c_1A^4+dF^2=DB^2\,,
\ee
\be
-d(1+D)F^2+b_1 DA^2-2a_1 D=2D(D-1-m)B^2\,,
\ee
\be
a_1 D+dF^2=(3m-D-Dm)B^2\,,
\ee
\be
a_2+dA^2-b_2F^2+c_2F^4=-(1-D)B^2\,,
\ee
\be
-dDA^2+b_2(1+D) F^2-2c_2F^4-2a_2 D=2(1-D)(m-D)B^2\,,
\ee
\be
a_2 D^2-b_2DF^2+c_2F^4=mD(1-D)B^2\,.
\ee

On solving these equations, we obtain the four unknowns $A,F,B,D$ as well
as two constraints between the seven parameters. We find
\bea\label{w.19}
&&A^2=\frac{2B^2[b_2(D^2-2D-2mD+3m)-d(m+2mD-2D-D^2)]}
{D(b_1b_2-d^2)}\,, \nonumber \\
&&F^2=\frac{2B^2[b_1(m+2mD-2D-D^2)-d(D^2-2D-2mD+3m)]}
{(1-D)(b_1b_2-d^2)}\,, \nonumber \\
&&B^2=\frac{3D(1-D)(m-D)(b_1b_2-d^2)^2}
{4c_1[b_2(D^2-2D-2mD+3m)-d(m+2mD-2D-D^2)]^2}\,, \nonumber \\
&&a_1D+dF^2=[3m-(1+m)D]B^2\,,
\eea
while the two constraints are
\be
D^2c_1A^4=(1-D)^2 c_2F^4\,,
~~a_2D-b_2F^2=\frac{B^2[D(mD+3)-2m(1+D)]}{(1-D)}\,.
\ee

At $m=1$ the solution reduces to the dark-dark soliton solution
\be\label{w.21}
\phi=\frac{A}{\sqrt{1-D\tanh^2(Bx+x_0)}}\,,~~
\psi=\frac{F\sech(Bx+x_0)}{\sqrt{1-D\tanh^2(Bx+x_0)}}\,,
\ee
provided
\bea\label{w.22}
&&A^2=\frac{2B^2(1-D)[b_2(3-D)-d(3D-1)]}
{D(b_1b_2-d^2)}\,, \nonumber \\
&&F^2=\frac{2B^2[b_1(1+D)-d(3-D)]}
{(b_1b_2-d^2)}\,, \nonumber \\
&&B^2=\frac{3D(b_1b_2-d^2)^2}
{4c_1[b_2(3-D)-d(1+D)]^2}\,,
~~a_1D+dF^2=[3m-(1+m)D]B^2\,,
\eea
while the two constraints are
\be\label{w.23}
D^2c_1A^4=(1-D)^2 c_2F^4\,,
~~a_2D-b_2F^2+(2+D)B^2=0\,.
\ee

{\bf Special case of $b_1b_2=d^2$}

One can show that the solution (\ref{w.18}) exists even in the case
$b_1b_2=d^2$. It turns out that such a solution exists only if
\be
\sqrt{b_2}\big[D^2-2(1+m)D+3m\big]=\sqrt{b_1}\big[m(1+2D)-D(2+D)\big]\,.
\ee
At $m=1$ this implies
\be\label{w.23}
\sqrt{b_2}(3-D)=\sqrt{b_1}(1+D)\,.
\ee

{\bf Energy}: Corresponding to the periodic solution (\ref{w.18}), 
the energy $\hat{E}$ and the constant $C$ are given by (using 
appropriate integrals in \cite{bf}) 
\bea\label{w.26}
&&\hat{E}=(A^2I_4+F^2I_2)B\,, \nonumber \\
&&C=\frac{m}{2}A^2B^2-\frac{F^2}{4D}\left[\frac{B^2(1-mD)}{(1-D)} 
-a_2\right]\,,
\eea
where $I_2$ and $I_4$ are given by Eqs. (\ref{2.40}), (\ref{2.41}),
(\ref{u.26a}) and (\ref{u.26b}). 

On using the expansion formulas for $E(k),K(k)$ and $\Pi(D,k)$ around
$k=1$ as derived above,
the energy of the periodic solution can be rewritten 
as the energy of the corresponding hyperbolic (dark-dark) soliton 
solution [Eq. (\ref{w.21})]  
plus the interaction energy. We find
\be\label{w.29}
\hat{E}=E_{kink}+E_{int}=BA^2[I_4^{(0)}+k'^2I_4^{(1)}]
+BF^2[I_2^{(0)}+k'^2I_2^{(1)}]\,,
\ee
where $I_{2,4}^{(0,1)}$ are given by Eqs. (\ref{kp9}), (\ref{kp10}),
(\ref{kp12}) and (\ref{kp13}).

Note that this solution exists only when $b_1 b_2 \ge d^2$.  The 
interaction energy vanishes at $k=1$.

\subsection{Solution XX}

Finally, yet another novel mixed phase solution is given by
\be\label{y.18}
\phi=\frac{A}{\sqrt{1-D\sn^2(Bx+x_0,m)}}\,,~~
\psi=\frac{F\dn(Bx+x_0,m)}{\sqrt{1-D\sn^2(Bx+x_0,m)}}\,.
\ee
This is a solution to the field Eqs. (\ref{2.2}) 
provided the following six equations are satisfied
\be
a_1-b_1A^2+c_1A^4+dF^2=DB^2\,,
\ee
\be
-d(m+D)F^2+b_1 DA^2-2a_1 D=2D(D-1-m)B^2\,,
\ee
\be
a_1 D+dmF^2=(3m-D-Dm)B^2\,,
\ee
\be
a_2+dA^2-b_2F^2+c_2F^4=-(m-D)B^2\,,
\ee
\be
-dDA^2+b_2(m+D) F^2-2mc_2F^4-2a_2 D=2(1-D)(m-D)B^2\,,
\ee
\be
a_2 D^2-b_2mDF^2+c_2m^2F^4=D(m-D)B^2\,.
\ee

On solving these equations, we obtain the four unknowns $A,F,B,D$ as well
as two constraints between the seven parameters. We find
\bea\label{y.19}
&&A^2=\frac{2B^2[b_2(D^2-2D-2mD+3m)-d(m+2D-2mD-D^2)]}
{D(b_1b_2-d^2)}\,, \nonumber \\
&&F^2=\frac{2B^2[b_1(m+2D-2mD-D^2)-d(D^2-2D-2mD+3m)]}
{(m-D)(b_1b_2-d^2)}\,, \nonumber \\
&&B^2=\frac{3D(1-D)(m-D)(b_1b_2-d^2)^2}
{(D^2-2D-2mD+3m)-d(m+2D-2mD-D^2)]^2}\,, \nonumber \\
&&a_1D+mdF^2=[3m-(1+m)D]B^2\,,
\eea
while the two constraints are
\be
D^2c_1A^4=(m-D)^2 c_2F^4\,,~~
a_2D-mb_2F^2=\frac{B^2[D^2+mD(3m-2)-2m^2]}{B^2}\,.
\ee

At $m=1$ the solution reduces to the dark-dark soliton solution
(\ref{w.21}) satisfying the relations (\ref{w.22}) and (\ref{w.23}). 

{\bf Special case of $b_1b_2=d^2$}

One can show that the solution (\ref{y.18}) exists even in the case
$b_1b_2=d^2$. It turns out that such a solution exists only if
\be
\sqrt{b_2}\big[D^2-2(1+m)D+3m\big]=\sqrt{b_1}\big[m+2D-2mD-D^2\big]\,.
\ee
At $m=1$ this implies the constraint (\ref{w.23}).

{\bf Energy}: Corresponding to the periodic solution (\ref{y.18}), 
the energy $\hat{E}$ and the constant $C$ are given by (using 
appropriate integrals in \cite{bf})  
\bea\label{y.26}
&&\hat{E}=(A^2I_4+F^2I_3)B\,, \nonumber \\
&&C=\frac{m}{2}A^2B^2-\frac{F^2}{4D}\left[\frac{mB^2(2m-m^2-D)}{(m-D)} 
-a_2\right]\,,
\eea
where $I_3$ and $I_4$ are given by Eqs. (\ref{2.49}) and (\ref{u.26a}), 
respectively.

On using the expansion formulas for $E(k),K(k)$ and $\Pi(D,k)$ around
$k=1$ derived above,
the energy of the periodic solution can be rewritten 
as the energy of the corresponding hyperbolic (dark-dark) soliton 
solution [Eq. (\ref{w.21})]  
plus the interaction energy. We find
\be\label{y.29}
\hat{E}=E_{kink}+E_{int}=BA^2[I_4^{(0)}+k'^2I_4^{(1)}]
+BF^2[I_3^{(0)}+k'^2I_3^{(1)}]\,,
\ee
where $I_3^{(0)}=I_2^{(0)}, I_3^{(1)}, I_4^{(0,1)}$ are  given by 
Eqs. (\ref{kp9}), (\ref{kp11}), (\ref{kp12}) and (\ref{kp13}).

Note that this solution exists only when $b_1 b_2 \ge d^2$.  The 
interaction energy vanishes at $k=1$, as it should.

{\bf Interaction energy:} Summarizing, we thus have obtained twenty 
solutions of the coupled $\phi^6$ model. We have calculated the 
corresponding soliton and interaction energy in each case. If we 
carefully look at the interaction energy expression in each case, we 
find that except for the solution XI, for all other solutions, the 
soliton interaction energy (in the dilute gas or asymptotic limit 
\cite{sanati}) has the leading term of the form $k'^2\ln(4/k')$. Out 
of these nineteen solutions, solutions I, II, III, XII and XV have 
repulsive interaction energy while the solutions IV, V, VI , XIII, 
XVI and XIX have attractive interaction energy. However, for the 
solutions VII, VIII, IX, X, XIV and XX, the interaction energy can 
be attractive or repulsive, depending on the values of the various 
parameters of the model. Finally, for solution XI, the interaction 
energy (in the dilute gas or asymptotic limit) has the leading term 
of the form $k'^2$ and is repulsive.  Since the lattice periodicity 
$r=2nK(k)/B$ with $n=1$ for $\dn$ based solutions, and $n=2$ for 
$\sn$ and $\cn$ based solutions, using $K(k)\sim \ln(4/k')$ we find 
that $k'^2=4n^2\exp(-rB/n)$.  Thus the interaction is purely 
exponential for solution XI whereas it is of the form $k'^2\ln(4/k') 
=(2rnB)\exp(-rB/n)$ for all other solutions.   

\section{Solutions of a Coupled $\phi^6$ Discrete Model}

We shall now consider a discrete variant of the above continuum coupled
$\phi^6$ model and obtain its static solutions.  We start from the 
coupled static field Eqs. (\ref{2.2}). The discrete analogue of these 
field equations has the form
\bea
&&(1/h^2)[\phi_{n+1}+\phi_{n-1}-2\phi_n]
=a_1\phi_n-b_1\phi_n^3+c_1\phi_n^5
+d\psi_n^2 \phi_n , \nonumber \\
&&(1/h^2)[\psi_{n+1}+\psi_{n-1}-2\psi_n]
=a_2\psi_n-b_2\psi_n^3+c_1\psi_n^5
+d\phi_n^2 \psi_n , \nonumber \\
\eea
where $h$ is the lattice spacing.  We are unable to find any solution of 
these coupled equations. However, as in the Ablowitz-Ladik discretization 
of the nonlinear Schr\"odinger equation \cite{al}, if we replace $\phi,\psi$ 
in the $\phi^5,\psi^5$ terms by the corresponding average, then we can 
find the exact solutions. Thus, we consider the coupled equations
\bea\label{3.1}
&&(1/h^2)[\phi_{n+1}+\phi_{n-1}-2\phi_n]
=a_1\phi_n-b_1\phi_n^3+\frac{c_1}{2}\phi_n^4[\phi_{n+1}+\phi_{n-1}]
+d\psi_n^2 \phi_n\,, \nonumber \\
&&(1/h^2)[\psi_{n+1}+\psi_{n-1}-2\psi_n]
=a_2\psi_n-b_2\psi_n^3+\frac{c_2}{2}\psi_n^4[\psi_{n+1}+\psi_{n-1}]
+d\phi_n^2 \psi_n\,. \nonumber \\
\eea
We shall show that while in the continuum case there are twenty periodic
solutions, we are able to obtain only six periodic solutions in the 
discrete case. We shall also see that none of the discrete solutions has 
a smooth continuum limit, i.e. in the limit $h \rightarrow 0$, none of 
them goes over to the continuum solutions obtained in the last section.

\subsection{Solution I}

It is easy to show that 
\be\label{3.2}
\phi_n=A\sqrt{m}\sn(hB[n+c],m)\,, ~~~ \psi_n=F\sqrt{m}\sn(hB[n+c],m)\,,
\ee
is a solution to the above equations provided 
\bea\label{3.3}
&&a_1=a_2=(2/h^2)[\cn(hB,m)\dn(hB,m)-1] <0\,,
~~(d+b_1)A^2=(d+b_2)F^2\,, \nonumber \\
&&A^2=\frac{2(d+b_2)\sn^2(hB,m)\cn(hB,m)\dn(hB,m)}
{h^2(d^2-b_1b_2)}\,, \nonumber \\
&&c_1A^4=c_2F^4\,,~c_1a_1
=\frac{[\cn(hB,m)\dn(hB,m)-1]
(d^2-b_1b_2)^2}{\cn^2(hB,m)\dn^2(hB,m)(b_2+d)^2}\,.
\eea
It is worth noting the completely different form of the 
solutions in the continuum and the discrete cases. We shall see that this
is true for all the six solutions that we obtain in the discrete case. 
Also note that while in
the continuum case $b_1b_2>d^2$, in the discrete case $d^2>b_1b_2$.

{\bf $m=1$ limit}

In the $m=1$ limit, this solution goes over to the bright-bright 
topological solution
\be\label{7.3}
\phi=A\tanh(hB[n+c])\,, ~~~ \psi=F\tanh(hB[n+c])\,,
\ee
provided
\bea\label{7.4}
&&a_1=a_2=-(2/h^2)\tanh^2(hB)<0\,,
~~A^2=\frac{2(d+b_2)\tanh^2(hB)\sech^2(hB)}
{h^2(d^2-b_1b_2)}\,, \nonumber \\
&&c_1a_1=-\frac{\tanh^2(hB)
(d^2-b_1b_2)^2}{(b_2+d)^2\sech^4(hB,m)}\,, 
\eea
while the other two relations are as given by Eq. (\ref{3.3}).
We note that this solution exists only if $d^2 > b_1 b_2$. 

In the limit $d=0$ (i.e. uncoupled case), the solution is given by
\be\label{p1}
\phi_n=A\sqrt{m}\sn(hB[n+c],m)\,,
\ee
provided
\bea\label{p2}
&&a_1=(2/h^2)[\cn(hB,m)\dn(hB,m)-1] <0\,,
~~A^2=-\frac{2\sn^2(hB,m)\cn(hB,m)\dn(hB,m)}
{h^2b_1}\,, \nonumber \\
&&\frac{c_1a_1}{b_1^2}
=\frac{[\cn(hB,m)\dn(hB,m)-1]}{\cn^2(hB,m)\dn^2(hB,m)}\,.
\eea

In the limit $m=1$ the uncoupled solution is given by
\be\label{p3}
\phi_n=A\tanh(hB[n+c])\,,
\ee
provided
\be\label{p4}
a_1=-\frac{2\tanh^2(hB)}{h^2}\,,
~~A^2=-\frac{2\tanh^2(hB)\sech^2(hB)}{h^2b_1}\,,
~~\frac{c_1a_1}{b_1^2}=-\frac{\tanh^2(hB)}{\sech^4(hB)}\,.
\ee
Thus, such a solution exists provided $a_1,b_1<0$.

\subsection{Solution II}

We shall now obtain three periodic solutions, all of which in the limit
$m \rightarrow 1$ go over to the same dark-dark soliton solution. The
first one of these, i.e.
\be\label{7.5}
\phi=A\sqrt{m}\cn(hB[n+c],m)\,, ~~~ \psi=F\sqrt{m}\cn(hB[n+c],m)\,,
\ee
is a solution to the above equations provided 
\bea\label{7.6}
&&a_1=a_2=(2/h^2)\left[\frac{\cn(hB,m)}{\dn^2(hB,m)}-1\right]\,,
~~A^2=\frac{2(d+b_2)\sn^2(hB,m)\cn(hB,m)}
{h^2(b_1b_2-d^2)\dn^4(hB,m)}\,, \nonumber \\
&&(d+b_1)A^2=(d+b_2)F^2\,,~~
c_1A^4=c_2F^4\,, \nonumber \\
&&c_1a_1=\left[\frac{\cn(hB,m)}{\dn^2(hB,m)}-1\right]
\frac{\dn^4(hB,m)(b_1b_2-d^2)^2}{\cn^2(hB,m)(b_2+d)^2}\,.
\eea
Again notice the completely different form of the 
solutions in the continuum and the discrete cases. One can show that
$a_1,a_2 > (<)\, 0$ provided $m> (<)\, 1/2$. Note that this solution
exists only if $b_1 b_2 > d^2$.

In the $m=1$ limit, three of the solutions (i.e. solutions II,
as well as solutions III and IV to be discussed below)
go over to the nontopological soliton solution
\be\label{q1}
\phi_n=A\sech(hB[n+c])\,, ~~~ \psi_n=F\sech(hB[n+c])\,,
\ee
provided
\bea\label{q2}
&&a_1=a_2=(2/h^2)[\cosh(hB)-1]>0\,,
~~A^2=\frac{2(d+b_2)\sinh^2(hB)\cosh(hB)}
{h^2(d^2-b_1b_2)}\,, \nonumber \\
&&(d+b_1)A^2=(d+b_2)F^2\,,~~
c_1A^4=c_2F^4\,, \nonumber \\
&&c_1a_1=[\cosh(hB)-1]
\frac{\sech^2(hB)(b_1b_2-d^2)^2}{(b_2+d)^2}\,.
\eea

In the limit $d=0$ (i.e. the uncoupled case), the solution is given by
\be\label{p5}
\phi_n=A\sqrt{m}\cn(hB[n+c],m)\,,
\ee
provided
\bea\label{p6}
&&a_1=(2/h^2)\left[\frac{\cn(hB,m)}{\dn^2(hB,m)}-1\right]\,,
~~A^2=\frac{2\sn^2(hB,m)\cn(hB,m)}
{h^2b_1\dn^4(hB,m)}\,, \nonumber \\
&&\frac{c_1a_1}{b_1^2}=\left[\frac{\cn(hB,m)}{\dn^2(hB,m)}-1\right]
\frac{\dn^4(hB,m)}{\cn^2(hB,m)}\,.
\eea

In the limit $m=1$, the uncoupled solution takes the form
\be\label{p7}
\phi_n=A\sech(hB[n+c])\,,
\ee
provided
\bea\label{p8}
&&a_1=(2/h^2)\left[\cosh(hB)-1\right]\,,
~~A^2=\frac{2\tanh^2(hB)}
{h^2b_1\sech^3(hB)}\,, \nonumber \\
&&\frac{c_1a_1}{b_1^2}=\left([1-\sech(hB)]\sech(hB)\right)\,.
\eea
Thus, the uncoupled solution exists only if $a_1,b_1>0$.

\subsection{Solution III}

It is easy to show that another periodic solution is given by
\be\label{7.7}
\phi=A\dn(hB[n+c],m)\,,~~~\psi=F\dn(hF[n+c],m)\,,
\ee
provided 
\bea\label{7.8}
&&a_1=a_2=(2/h^2)\left[\frac{\dn(hB,m)}{\cn^2(hB,m)}-1\right]>0\,,
~~A^2=\frac{2(d+b_2)\sn^2(hB,m)\dn(hB,m)}
{h^2(b_1b_2-d^2)\cn^4(hB,m)}\,, \nonumber \\
&&(d+b_1)A^2=(d+b_2)F^2\,,~~
c_1A^4=c_2F^4\,, \nonumber \\
&&c_1a_1=\left[\frac{\dn(hB,m)}{\cn^2(hB,m)}-1\right]
\frac{\cn^4(hB,m)(b_1b_2-d^2)^2}{\dn^2(hB,m)(b_2+d)^2}\,.
\eea
Note that as in the continuum case, here too $a_1,a_2>0$.
Further, this solution exists only if $b_1 b_2 > d^2$.
Finally, in the limit $m=1$, this solution goes over to the solution
(\ref{q1}) satisfying the constraints (\ref{q2}).

In the limit $d=0$ (i.e. the uncoupled case), the solution is given by
\be\label{p9}
\phi=A\dn(hB[n+c],m)\,,
\ee
provided
\bea\label{p10}
&&a_1=(2/h^2)\left[\frac{\dn(hB,m)}{\cn^2(hB,m)}-1\right]\,,
~~A^2=\frac{2\sn^2(hB,m)\dn(hB,m)}
{h^2b_1\cn^4(hB,m)}\,, \nonumber \\
&&\frac{c_1a_1}{b_1^2}=\left[\frac{\dn(hB,m)}{\cn^2(hB,m)}-1\right]
\frac{\cn^4(hB,m)}{\dn^2(hB,m)}\,.
\eea

In the limit $m=1$, the uncoupled solution goes over to the solution
(\ref{p7}) satisfying relations (\ref{p8}).

\subsection{Solution IV}

It is easy to show that yet another periodic solution is
given by
\be\label{7.9}
\phi=A\sqrt{m}\cn(hB[n+c],m)\,, ~~~ \psi=F\dn(hB[n+c],m)\,,
\ee
provided 
\bea
&&a_1=(2/h^2)\left[\frac{\cn(hB,m)}{\dn^2(hB,m)}-1\right]\,,
a_2=(2/h^2)\left[\frac{\dn(hB,m)}{\cn^2(hB,m)}-1\right]>0\,, \nonumber \\
&&A^2=\frac{2[d\dn^{5}(hB,m)+b_2\cn^{5}(hB,m)]\sn^2(hB,m)}
{h^2(b_1b_2-d^2)\cn^4(hB,m)\dn^{4}(hB,m)}\,, \nonumber \\
&&F^2=\frac{2[b_1\dn^{5}(hB,m)+d\cn^{5}(hB,m)]\sn^2(hB,m)}
{h^2(b_1b_2-d^2)\cn^4(hB,m)\dn^{4}(hB,m)}\,, \nonumber \\
&&c_1A^4=\frac{2\sn^4(hB,m)}{h^2\dn^4(hB,m)}\,,~~
c_2F^4=\frac{2\sn^4(hB,m)}{h^2\cn^4(hB,m)}\,.
\eea
Note that $a_2 >0$ while $a_1 > (<)\, 0$ depending on if $m > (<)\, 1/2$. 

Again note that like the previous two solutions, this one also exists 
only if $b_1 b_2 >d^2$.
As mentioned above, in the limit $m=1$, this solution also goes over to
the solution (\ref{q1}) satisfying the constraints (\ref{q2}). Further,
in the decoupled case (i.e. $d=0$), $\phi_n$ and $\psi_n$ satisfy solutions
(\ref{p5}) and (\ref{p9}), respectively, with the appropriate constraints.


\subsection{Solution V}

Finally, there are two periodic solutions, both of which at $m=1$ go over
to the (same) bright-dark soliton solution.
The first one is given by
\be
\phi=A\sqrt{m}\sn(hB[n+c],m)\,, ~~~ \psi=F\sqrt{m}\cn(hB[n+c],m)\,,
\ee
provided $d<0$ and further 
\bea
&&a_1-m|d|F^2=(2/h^2)[\cn(hB,m)\dn(hB,m)-1]<0\,,  \nonumber \\
&&a_2-m|d|A^2=(2/h^2)\left[\frac{\cn(hB,m)}{\dn^2(hB,m)}-1\right]\,, 
\nonumber \\
&&A^2=\frac{2[|d|-b_2\dn^{5}(hB,m)]\sn^2(hB,m)\cn(hB,m)}
{h^2(b_1b_2-d^2)\dn^{4}(hB,m)}\,, \nonumber \\
&&F^2=\frac{2[b_1-|d|\dn^{5}(h\beta,m)]\sn^2(h\beta,m)\cn(h\beta,m)}
{h^2(b_1b_2-d^2)\dn^{4}(hB,m)}\,, \nonumber \\
&&c_1A^4=\frac{2\sn^4(hB,m)}{h^2}\,,~~
c_2F^4=\frac{2\sn^4(hB,m)}{h^2\dn^4(hB,m)}\,.
\eea

Note that this solution exists only if $b_1 b_2 >d^2$.  In the decoupled 
limit (i.e. $d=0$), the $\phi_n$ and $\psi_n$ fields go over to the 
solutions (\ref{p1}) and (\ref{p5}), respectively, satisfying appropriate 
constraints.

\subsection{Solution VI}

Finally, the second solution is given by
\be
\phi=A\sqrt{m}\sn(hB[n+c],m)\,, ~~~ \psi=F\dn(hB[n+c],m)\,,
\ee
provided $d<0$ and further 
\bea
&&a_1-|d|F^2=(2/h^2)[\cn(hB,m)\dn(hB,m)-1]<0\,, \nonumber \\
&&a_2-|d|A^2=(2/h^2)\left[\frac{\dn(hB,m)}{\cn^2(hB,m)}-1\right]>0\,, 
\nonumber \\
&&A^2=\frac{2[|d|-b_2\cn^{5}(hB,m)]\sn^2(hB,m)\dn(hB,m)}
{h^2(b_1b_2-d^2)\cn^4(hB,m)}\,, \nonumber \\
&&F^2=\frac{2[b_1-|d|\cn^{5}(hB,m)]\sn^2(hB,m)\dn(hB,m)}
{h^2(b_1b_2-d^2)\cn^4(hB,m)}\,, \nonumber \\
&&c_1A^4=\frac{2\sn^4(hB,m)}{h^2}\,,~~
c_2F^4=\frac{2\sn^4(hB,m)}{h^2\cn^4(hB,m)}\,.
\eea
Note that $a_2 >0$. 

In the limit $m=1$, both of these solutions (i.e. solutions V and VI) go over
to the bright-dark soliton solution
\be
\phi=A\tanh(hB[n+c])\,, ~~~ \psi=F\sech(hB[n+c])\,,
\ee
provided $d<0$ and further 
\bea
&&a_1-|d|F^2=-(2/h^2)\tanh^2(hB) \,,
a_2-|d|A^2=(2/h^2)[\cosh(hB)-1]\,, \nonumber \\
&&A^2=\frac{2[|d|-b_2\sech^{5}(hB)]\tanh^2(hB)}
{h^2(b_1b_2-d^2)\sech^3(hB,m)}\,, \nonumber \\
&&F^2=\frac{2[b_1-|d|\sech^{5}(hB)]\tanh^2(hB)}
{h^2(b_1b_2-d^2)\sech^3(hB)}\,, \nonumber \\
&&c_1A^4=\frac{2\tanh^4(hB)}{h^2}\,,~~
c_2F^4=\frac{2\sin^4(hB)}{h^2}\,.
\eea
Note that this solution exists only if $b_1 b_2 >d^2$.  In the decoupled 
limit (i.e. $d=0$), the $\phi_n$ and the $\psi_n$ fields go over to the 
solutions (\ref{p1}) and (\ref{p9}), respectively, satisfying appropriate 
constraints.

As in the continuum case, can one say if these discrete solutions are
valid below, above or at $T_c$?  There is a note of caution here. Since 
we are considering a discrete model here, and since the corresponding 
Hamiltonian from which the coupled discrete field Eqs. (\ref{3.1}) can 
be derived is not known, strictly speaking we cannot directly associate 
$a_1$, $b_1$ and other parameters of the discrete problem and draw 
conclusions about the nature of the transition by using the corresponding 
continuum model potential (\ref{2.1}).

However, as a first guess, it might be reasonable to assume that for 
the same values of the ratio of the parameters (i.e. $a_1c_1/b_1^2$)
as in the continuum case, one can classify if the solution is below,
above or at $T_c$. For example, since solution I exists if $a_1,b_1<0$ 
($c_1$ is always assumed to be positive), this implies that this 
solution exists with $T < T_c^{II}$.  It is worth reminding that in 
case $b_1<0$, then the uncoupled $\phi^6$ model characterized by the 
potential (\ref{2.1}) corresponds to a second order (and not a first 
order) transition.  Similarly, since for solutions II, III and IV, 
$b_1,a_1>0$ and $b_1^2\ge 4a_1c_1$, this implies that $T < T_p^{I}$. 
Finally, solutions V and VI correspond to mixed-type solutions with 
$\phi_n$ field corresponds to $T < T_c^{II}$ while $\psi_n$ field 
corresponds to $T < T_p^{I}$.

Since we do not know the Hamiltonian associated with Eqs. (\ref{3.1}), 
we are unable to find the soliton and interaction energy explicitly 
for these discrete solutions and hence cannot comment whether the 
Peierls-Nabarro barrier is zero or non-zero. 
 
\section{Conclusion}

We have provided a set of twenty distinct exact, periodic domain wall 
solutions for a coupled $\phi^6$ model with biquadratic coupling. These
in turn lead to nine distinct (hyperbolic) soliton solutions both at, above 
and below $T_c$. The corresponding discrete case was also considered and we 
found six different periodic solutions which in turn lead to three distinct
(hyperbolic) soliton solutions. The soliton and interaction energy were 
obtained for all twenty continuum solutions.  For the six solutions of the 
discrete model, the calculation of the Peierls-Nabarro barrier \cite{PN1,PN2,pgk} 
and soliton scattering \cite{panos} are important topics of further study.  
Similarly, scattering of solitons for the twenty solutions in the coupled 
$\phi^6$ continuum model is an interesting issue with these static solutions 
boosted with a certain velocity.  We also discovered previously unknown 
periodic solutions of the uncoupled $\phi^6$ model \cite{behera,falk,sanati}.  

It would be useful to explore whether the different solutions are completely 
disjoint or if there are any possible bifurcations linking them via, for 
instance, analytical continuation.  We have not carried out an explicit 
stability analysis of various periodic solutions \cite{zim}.  However,
the energy calculations and interaction energy between solitons (for
$m\sim1$) could provide useful insight in this direction.  Similarly, it 
would be worth exploring the problem of a coupled $\phi^6$ model in the 
presence of an external field.  Our results are relevant for domain walls 
in structural phase transitions in ferroelectrics and elastic materials 
\cite{sbh,hs,hlss,falk,sanati,das} and possibly in field theoretic 
contexts \cite{bazeia,santos,lou}. These ideas and solutions can be 
generalized to other coupled models for first order transitions such as 
coupled asymmetric double well potentials ($\phi^2$-$\phi^3$-$\phi^4$) 
and will be discussed elsewhere.

\section{Acknowledgment}
A.K. acknowledges the hospitality of the Center for Nonlinear
studies at LANL.  This work was supported in part by the U.S.
Department of Energy.

\end{document}